\documentclass{aa}  

\usepackage{graphicx}
\usepackage{txfonts}
\usepackage{xcolor}

\begin{document} 

   \title{A UNIONS view of the brightest central galaxies of candidate fossil groups~\thanks{Based on observations obtained with CFHT, SDSS, CAHA, and OHP observatories 
(see acknowledgements for more details).}}

   \author{A. Chu
          \inst{1}
          \and
          F. Durret
          \inst{1}
          \and
          A. Ellien
          \inst{2}
          \and
          F. Sarron
          \inst{3}
          \and
          C. Adami
          \inst{4}
          \and
         I. M\'arquez
          \inst{5}
          \and
          N. Martinet
          \inst{4}
          \and
          T. de Boer
          \inst{6}
          \and
          K. C. Chambers
          \inst{6}
          \and
          J.-C.~Cuillandre
          \inst{7}
          \and
          S.~Gwyn
          \inst{8}
          \and
          E. A. Magnier
          \inst{6}
          \and
          A.~W.~McConnachie
          \inst{8}
}

\institute{Sorbonne Universit\'e, CNRS, UMR 7095, Institut d'Astrophysique de Paris, 98bis Bd Arago, 75014, Paris, France
\and
Anton Pannekoek Institute for Astronomy \& GRAPPA, University of Amsterdam, Science Park 904, 1098 XH Amsterdam, The Netherlands 
\and
IRAP, Institut de Recherche en Astrophysique et Plan\'etologie, Universit\'e de Toulouse, UPS-OMP, CNRS, CNES, 14 avenue E.~Belin, F-31400 Toulouse, France
\and
Aix-Marseille Univ., CNRS, CNES, LAM, Marseille, France
\and
Instituto de Astrof\'isica de Andaluc\'ia, CSIC, Glorieta de la Astronom\'ia s/n, 18008, Granada, Spain
\and
Institute for Astronomy, University of Hawaii, 2680 Woodlawn Drive, Honolulu HI 96822, USA
\and
Universit\'e Paris-Saclay, Universit\'e Paris Cit\'e, CEA, CNRS, AIM, 91191, Gif-sur-Yvette, France 
\and
Herzberg Astronomy and Astrophysics, National Research Council, 5071 West Saanich Road, Victoria, BC, Canada, V9E2E7
}

\date{Received }

  \abstract
   {The formation process of fossil groups (FGs) is still under debate, and, due to the relative rarity of FGs, large samples of such objects are still missing. }
   {The aim of the present paper is to increase the sample of known FGs, and to analyse the properties of their brightest group galaxies (BGG) and compare them with a control sample of non-FG BGGs. }
   {Based on the large spectroscopic catalogue of haloes and galaxies publicly made available by Tinker, we extract a sample of 87 FG and 100 non-FG candidates. 
   For all the objects with data available in UNIONS  (initially the Canada France Imaging Survey, CFIS),
    in the u and r bands, and/or in an extra r-band processed to preserve all low surface brightness features (rLSB hereby), we made a 2D photometric fit of the BGG with GALFIT with one or two S\'ersic components. We also analysed how the subtraction of intracluster light contribution modifies the BGG properties. From the SDSS spectra available for the BGGs of 65 FGs and 82 non-FGs, 
   we extracted the properties of their stellar populations with Firefly. To complement our study, we investigated the origin of the emission lines in a nearby FG, dominated by the NGC~4104 galaxy, to illustrate in detail the possible origin of emission lines in the FG BGGs, involving the presence or absence of an AGN. 
   }
   {Morphologically, a single S\'ersic profile can fit most objects in the u band, while two S\'ersics are needed in the r and rLSB bands, both for FGs and non-FGs. Non-FG BGGs cover a larger range of S\'ersic index $n$. FG BGGs follow the Kormendy relation (mean surface brightness versus effective radius) previously derived for almost one thousand brightest cluster galaxies (BCGs) by \cite{Chu+22} while non-FGs BGGs are in majority located below this relation, with fainter mean surface brightnesses. This suggests that FG BGGs have evolved similarly to BCGs, and non-FG BGGs have evolved differently from both FG BGGs and BCGs. All the above properties can be strongly modified by the subtraction of intracluster light contribution. Based on spectral fitting, the stellar populations of FG and non-FG BGGs do not differ significantly.
   }
   {The morphological properties and the Kormendy relation of FG and non-FG BGGs differ, suggesting they have had different formation histories. However, it is not possible to trace differences in their stellar populations or in their large scale distributions.}

   \keywords{Galaxies: fossil groups, morphology. }

   \maketitle
%

\section{Introduction}

Fossil groups (FGs) were discovered by \cite{Ponman+94}. They are particular groups of galaxies with high X-ray luminosities but with fewer bright galaxies than groups or clusters of galaxies. 
\cite{Jones+03} later gave the commonly accepted definition of FGs as satisfying three conditions: they are extended X-ray sources with an X-ray luminosity of at least 
L$_{\rm X}$ = 10$^{42}$ h$^{-2}_{50}$ erg~s$^{-1}$, with a Brightest Group Galaxy (BGG) at least two magnitudes brighter than other group members, the distance between the two brightest galaxies being smaller than half the group virial radius. The formation of these peculiar objects and why they present such a low amount
of optically emitting matter are still under debate. \cite{Jones+03} have suggested that FGs are the remnants of early mergers, and that
they are cool-core structures which accreted most of the large galaxies in their environment a long time ago, a 
scenario supported by hydrodynamical simulations by \cite{DOnghia+05}. However, FGs could also
be a short temporary stage of group evolution before they capture more galaxies in their vicinity, as reported for instance
by \cite{vonBenda08}, based on N-body simulations.

FGs can be studied through their optical \citep{Vikhlinin+99,Santos+07} or X-ray \citep{Romer+00,Adami+18} properties. 
Some optical studies support the scenario that FGs are the result of a large dynamical activity at high redshift, but in
an environment that is too poor for them to evolve into a cluster of galaxies through the hierarchical growth of structures. 
For example, \cite{LaBarbera+09} found that the optical properties of BGGs in FGs are identical to those of giant isolated field galaxies.  
\cite{Girardi+14} found a similar relation between their X-ray and optical luminosities for FGs and for normal groups, 
suggesting that all groups contain the same amount of optical material, but that in FGs it is concentrated in a giant central elliptical galaxy that has cannibalized most of the surrounding bright galaxies. 
At X-ray wavelengths, based on Chandra X-ray observations, \cite{Bharadwaj+16} found that FGs are mostly cool-core systems, suggesting that these structures are no longer dynamically active.

However recent observations tend to contradict the findings that FGs are dynamically relaxed systems that have not undergone recent merging events. For example, \cite{Kim+18} reported that the prototypical FG NGC~1132 shows an asymmetrical disturbed X-ray profile, and suggested that it is dynamically active. Similarly, \cite{LimaNeto+20} detected shells around the BGG of NGC~4104 and, based on N-body simulations, showed that this FG has probably experienced a relatively recent merger between its BGG and another
bright galaxy with a mass of about 40\% of that of the BGG. More details on FGs can be found in the recent review by \cite{Aguerri+21}.

To make up for the lack of large samples of FGs, \cite{Adami+20} made a statistical study of FGs, extracted from the catalog of 1371 groups and clusters detected by \cite{Sarron+18} in the Canada-France-Hawaii Telescope Legacy Survey (CFHTLS). These systems were detected based on photometric redshifts \citep{Ilbert+06}. \cite{Adami+20} found that groups with masses larger than $2.4\times 10^{14}$~M$_\odot$ had the highest probability to be FGs and discussed their location in the cosmic web relatively to nodes and filaments \citep[for a similar study, see also][]{Zarattini+22}. They concluded that FGs were most probably in a poor environment making the number of nearby galaxies insufficient to compensate for the accretion by the central group galaxy.

Numerical simulations of FGs by \cite{Dariush+07} have shown a good agreement with both optical and X-ray observations, and predict that fossil systems will be found in significant numbers (3-4\% of the population), even for quite rich clusters.  
They find that FGs assemble a higher fraction of their mass at high redshifts than
non-fossil groups, with the ratio of the currently assembled halo mass to final mass, at any
epoch, being about 10–20\% higher for FGs. Their interpretation is that FGs represent undisturbed, early-forming systems in which large galaxies have merged to form a single dominant elliptical.

The role of the BGG is therefore crucial to explain the lack of bright galaxies in FGs. The aim of the present paper is to analyse the physical properties of the BGGs of FGs and compare them to those of non-fossil groups and clusters. For this, we gathered a sample of FGs as large as possible from the sample of groups detected by Tinker from SDSS data (Section~\ref{sec:data}). In each FG candidate, we detected the BGG and measured its morphological properties. We then compared the properties of FG BGGs to those of a control sample of non-FGs, as well as to the brightest galaxies of clusters and massive groups of galaxies previously studied by \cite{Chu+21,Chu+22} to study how FG BGGs compare to the BCGs and BGGs of more imposing systems (Section~\ref{sec:morpho}). We also analysed the stellar populations of FG and non-FG BGGs and
investigated the origin of the BGG spectroscopic emission lines, taking as an example a very nearby fossil group BGG NGC~4104
(Section~\ref{sec:popstell}).
Finally, our results are discussed in Section~\ref{sec:discu}.

\section{Data}
\label{sec:data}
\subsection{Selection of FGs} 

Tinker has made available 
catalogues\footnote{https://www.galaxygroupfinder.net/catalogs} with data for
559,038 galaxies. These catalogues give, among other quantities, positions, spectroscopic redshifts from the SDSS survey, k-corrected
and evolution corrected (to z=0.1) g and r band absolute magnitudes, galaxy
stellar masses, and total halo masses. For each galaxy, the group to which it belongs is indicated.

The group-finding algorithm described by \cite{Tinker21} is based on the halo-based group finder of \cite{Yang+05}, further vetted by \cite{Campbell+15}. The standard implementation of the group finder yields central galaxy samples with a purity and completeness of 85–90 per cent \citep{Tinker+11}.
To assign stellar masses to haloes and subhaloes, \cite{Tinker21}  uses the stellar mass function from \cite{Cao+20}, which utilizes the principal component analysis galaxy stellar masses of \cite{Chen+12}.

We first eliminated all the galaxies that were alone in a group, because a single galaxy in a halo does not form a group, and obtained a catalogue of
201,007 galaxies that were at least in a pair.
We then selected the galaxies belonging to groups
where the magnitude difference in the r band between the brightest and
second brightest galaxy was at least 2~magnitudes, and for which the distance between these two galaxies was smaller than half the virial radius, $r_{vir}$:
$$R_{virial} = ( M_{halo} \times 4.30091\times 10^{-9}) / (100 \times H(t)^2) ^{1/3},$$
\noindent
where the corresponding mass $M_{halo}$
is given in the Tinker catalogue, G is the gravitational constant and $H(t)$ the Hubble constant at the group redshift.
We thus obtained a catalogue of 2453 galaxies.
We note that in this process, we do not take into account the possibility to have a projected galaxy at less than half the virial radius, which could in fact physically (in 3D) be at more than half the virial radius (because of the lack of precision of the spectroscopic redshift or because of high proper velocity). Such a galaxy could artificially be placed in the group luminosity function between the BGG and the second ranked galaxy, and therefore unqualify the
group as a fossil group. However, such a case would just limit the size of our sample, wrongly eliminating some of the fossil groups. This effect will
not pollute our fossil group sample by inserting non fossil groups. 

We extracted from the above catalogue a list with the brightest galaxy of
each group, and this led to a catalog of 1112 galaxies that may be
considered as BGGs. This means that in fact most groups were made of pairs.
In order to avoid considering objects that could
not be real groups, such as isolated galaxies with a few small
satellites, we added a condition on the halo mass:
$M_{halo}>10^{13}$~M$_\odot$, giving 88 FG candidates. This limit was chosen to match the
lowest mass that we found for a FG in our search for FGs in the CFHTLS: $1.1\times 10^{13}$~M$_\odot$ \citep{Adami+20}. 
In absence of X-ray data for all our objects except one, this condition also gives more confidence that these systems may indeed be FGs.
Indeed, N. Clerc kindly matched our FG catalogue  with his XCLASS catalogue of X-ray sources derived from XMM-Newton data and only found one match for FG17 (e.g. FG \#17 of Table~\ref{tab:88FG}), with L$_{\rm X}$ = 5 $\times$ 10$^{41}$ h$^{-2}_{50}$ erg~s$^{-1}$ in the [0.5 - 2] keV energy band. This value is slightly lower than the limit of $10^{42}$ h$^{-2}_{50}$ erg~s$^{-1}$ defined by \cite{Jones+03}.
For the other FGs, this does not mean that they are not X-ray emitters, but simply that they are not located in regions observed by XMM-Newton.

The last step was to use photometric catalogs from UNIONS\footnote{https://www.skysurvey.cc}
to check if no galaxies were missed in spectroscopy. The Ultraviolet Near Infrared Optical Northern Survey (UNIONS) collaboration combines wide field imaging surveys of the northern hemisphere. UNIONS consists of the Canada-France Imaging Survey (CFIS), conducted at the 3.6-meter CFHT on Maunakea, parts of Pan-STARRS, and the Wide Imaging with Subaru HyperSuprime-Cam of the Euclid Sky (WISHES). CFHT/CFIS is obtaining deep u and r bands; Pan-STARRS is obtaining deep i and moderate-deep z band imaging, and Subaru is obtaining deep z-band imaging through WISHES and g-band imaging through the Waterloo-Hawaii IfA g-band Survey (WHIGS). These independent efforts are directed, in part, to securing optical imaging to complement the Euclid space mission, although UNIONS is a separate collaboration aimed at maximizing the science return of these large and deep surveys of the northern skies.

We searched in the UNIONS photometric archive all objects (galaxies or stars) that 1) fall within 0.5$\times$R$_{virial}$ of a FG center, 2) are missing from the Tinker catalog, and 3) have magnitudes between that of the BGG and that of the second brightest galaxy.
Stars were removed with central surface brightness versus total magnitude plots.

\subsection{Additional spectroscopic observations of FGs}

At the end of this selection process, we found three galaxies not present in the  Tinker spectroscopy and potentially contributing to the first two magnitude range. This affected two FG candidates: FG65 and FG73 (see Table~\ref{tab:88FG}). 

For FG65, we obtained long-slit spectroscopy with MISTRAL at Observatoire de Haute-Provence for the two relatively bright galaxies
(RA=228.7630093$^{\rm o}$, DEC=42.0503814$^{\rm o}$ and RA=228.769596$^{\rm o}$, DEC=42.0548771$^{\rm o}$, 1 hour 
exposures), which both have magnitudes differing by less than 2~magnitudes from the BGG. Both proved to be part of the same foreground galaxy at z=0.0149, and not related to the FG BGG at z=0.13479. This confirmed the fossil nature of the FG65 group.

We also observed a galaxy (RA=238.45581$^{\rm o}$, DEC=56.4229$^{\rm o}$, 1~hour exposure) within the FG73 FG candidate 
with CAFOS at the Calar Alto Hispano Alem\'an Telescope. This galaxy proved to be at a redshift of 0.1055, very close to the redshift of the putative BGG (z=0.1080), and to have less than two magnitudes difference with the BGG. FG73 was therefore removed from the FG final list because the two magnitude difference criterium between the brightest and second brightest galaxies was not satisfied.

We thus obtained the final catalogue of 87 FG candidates
listed in Table~\ref{tab:88FG}.

\subsection{Basic properties of FGs}

The histograms of the halo masses and BGG stellar masses of the FG candidates and of the comparison sample (see below) are shown in
Figs.~\ref{fig:histomasshalo88} and \ref{fig:histomass88} respectively. 
These quantities are taken from Tinker's catalogues.
The halo masses are in the [$10^{13}$,$10^{14}$]~M$_\odot$ range,
with 65 FGs (74\%) having halo masses $\leq 2\times 10^{13}$~M$_\odot$. The BGG stellar masses are in the [$10^{10.8}$,$10^{12}$]~M$_\odot$ range, except for one galaxy (FG22) that has a lower mass of $10^{10.4}$~M$_\odot$.

\begin{figure}[h]
\begin{center}
\includegraphics[width=0.49\textwidth]{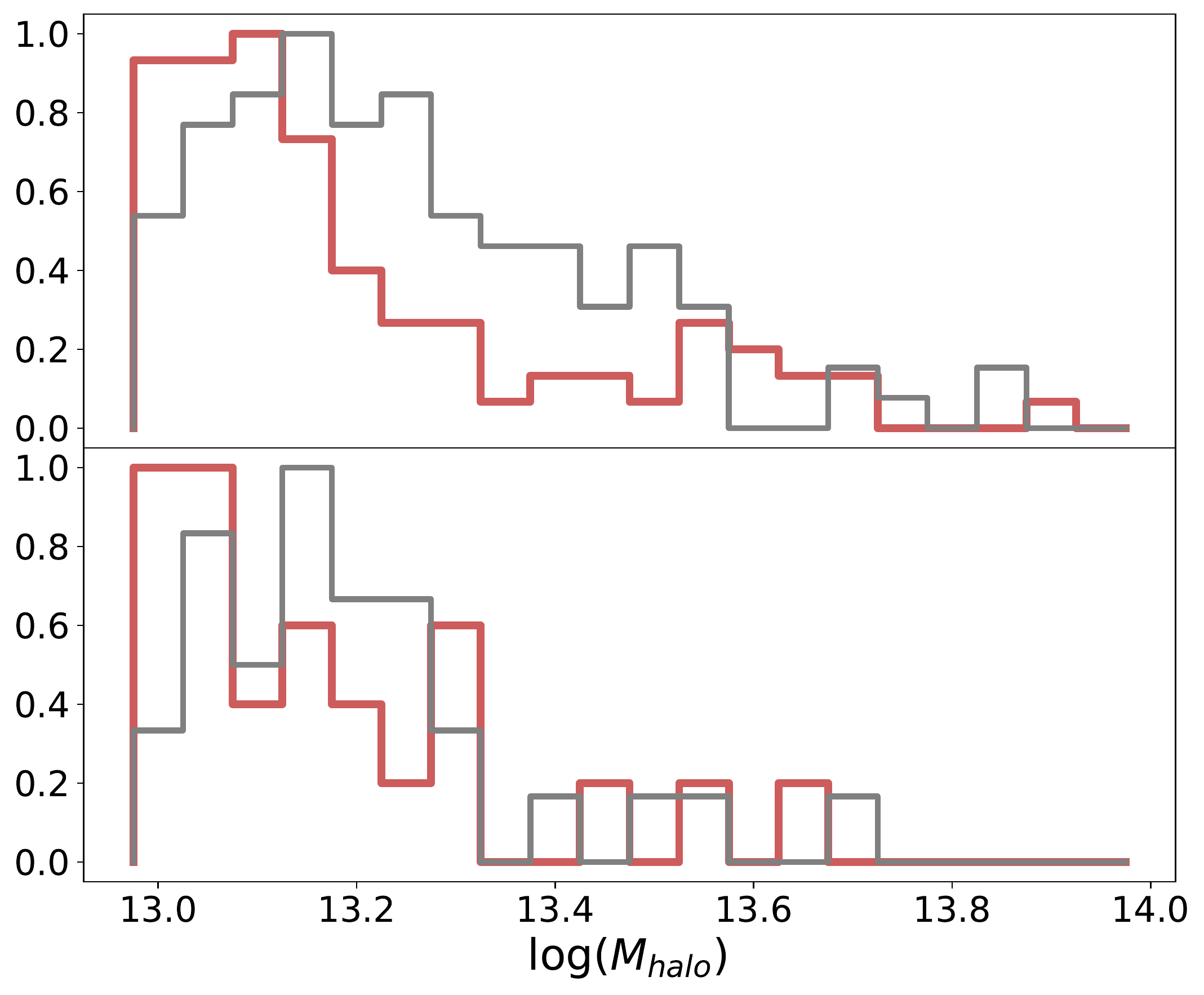}
\end{center}
\caption{Normalized histograms of the logarithms of the halo masses for
FGs (red) and non-FGs (grey). Top: the 87 FGs and 100 non-FGs. Bottom: 25 FGs and 30 non-FGs analysed morphologically.}
\label{fig:histomasshalo88}
\end{figure}

\begin{figure}[h]
\begin{center}
\includegraphics[width=0.49\textwidth]{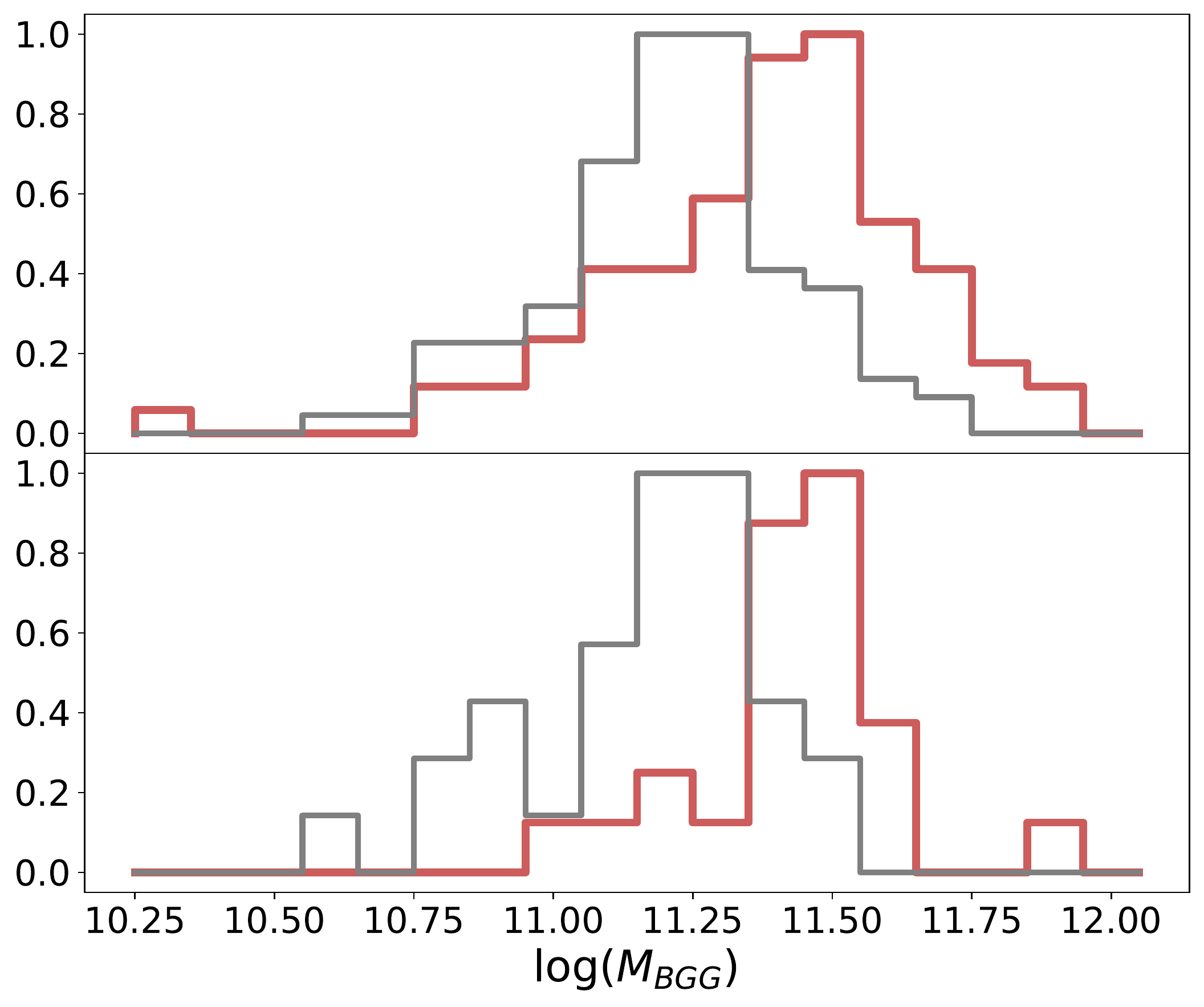}
\end{center}
\caption{Same as Fig.~\ref{fig:histomasshalo88} for the
normalized histograms of the logarithms of the BGG stellar masses.}
\label{fig:histomass88}
\end{figure}

\begin{figure}[h]
\begin{center}
\includegraphics[width=0.49\textwidth]{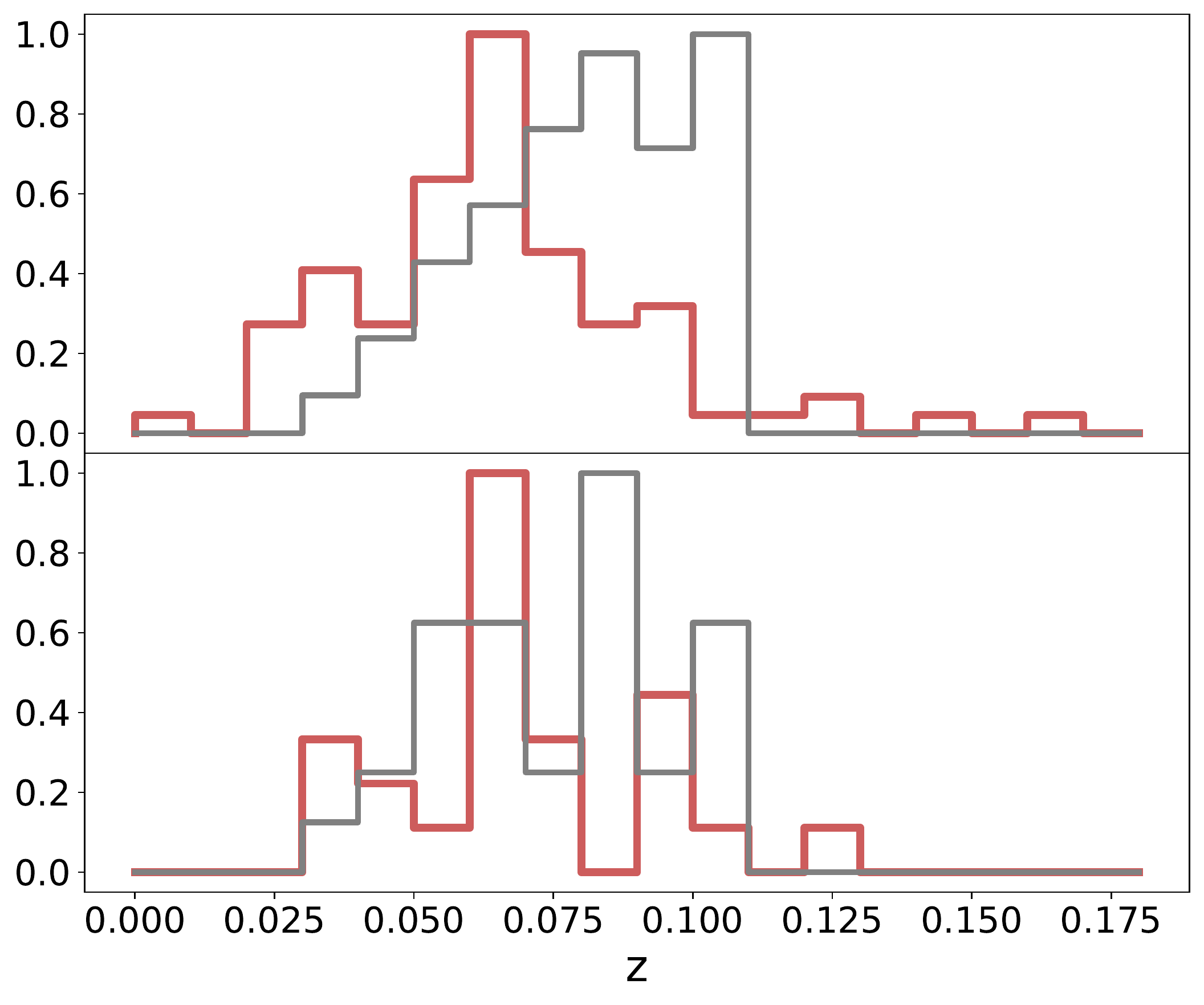}
\end{center}
\caption{Same as Fig.~\ref{fig:histomasshalo88 for the
normalized redshift histograms.}}
\label{fig:histoz88}
\end{figure}

\begin{figure}[h]
\begin{center}
\includegraphics[width=0.49\textwidth]{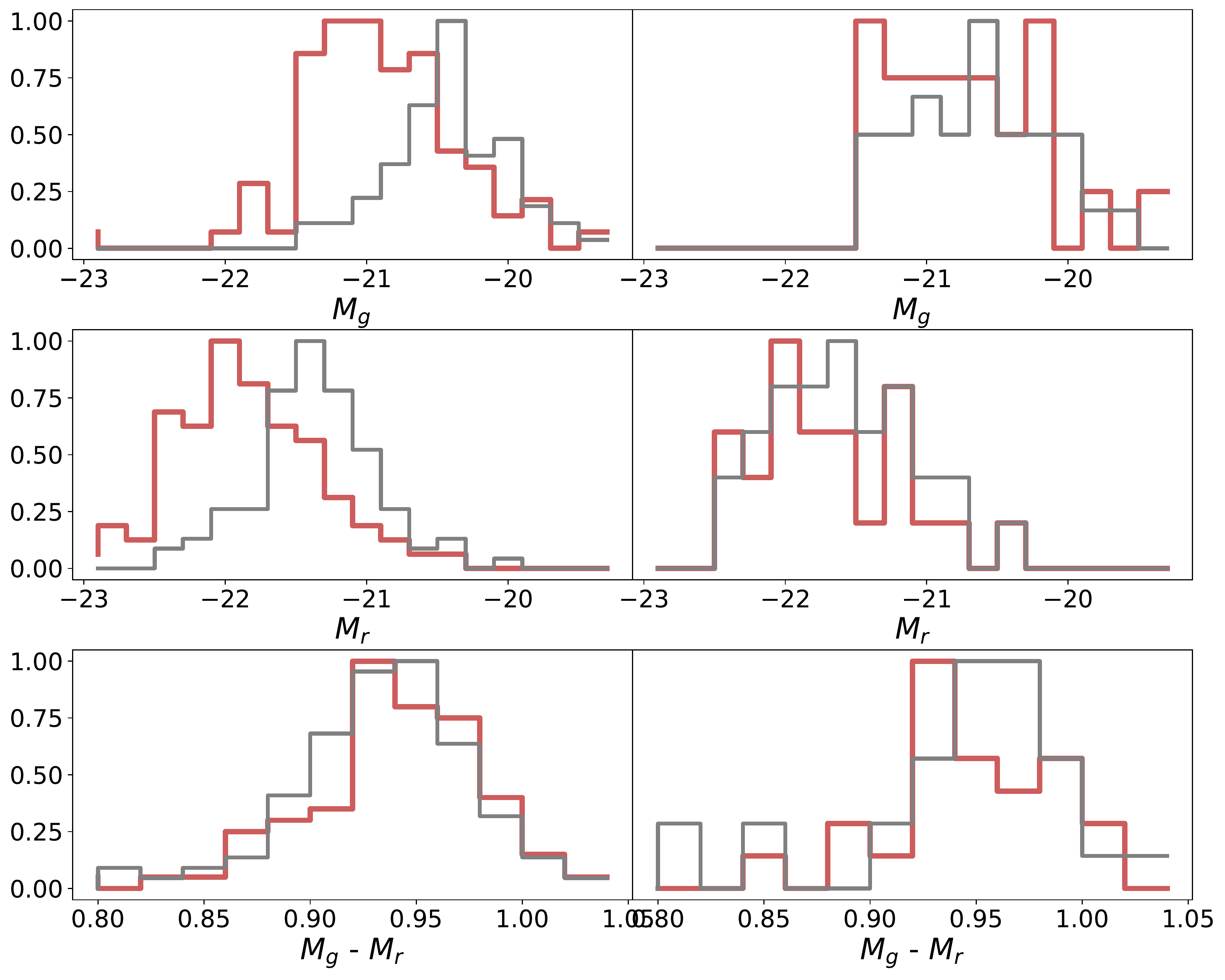}
\end{center}
\caption{Normalized histograms of the absolute magnitudes of the BGGs in the g (top) and r (middle) bands, and of the (g-r) colour (bottom), for FGs (red) and non-FGs (grey). Left: the 87 FGs and 100 non-FGs. Right: 25 FGs and 30 non-FGs analysed morphologically.}
\label{fig:histomags}
\end{figure}

The redshift histogram of the groups is shown in
Fig.~\ref{fig:histoz88}. Most of the FGs are in the [0.02,0.12] range, with only five FGs in the [0.12,0.18] range. The absolute magnitude histograms in the g and r bands, together with the (g-r) colour histogram, are shown in Fig.~\ref{fig:histomags} for the groups. BGGs have typical absolute magnitudes in the r band in the [$-23,-21$] range, with typical (g-r) colors in the [0.8,1.0] range.

We searched for images of these 87 FG candidates in the UNIONS image database in the u
and r bands. These images were retrieved from the DR3, which includes observations made before January of  2021, and covers 7000 deg$^2$ in u and 3600 deg$^2$ in r.  
We found images for 35 FGs in the u band and for 25 FGs in the r band, among which 12 FGs having both u and r band images.
We indicate in Table~\ref{tab:88FG} (last column) the FGs with UNIONS images available.

\cite{Chu+22} have shown that subtracting the contribution of intracluster light (ICL) could modify the properties derived for BCGs. Therefore, with the aim of estimating how subtracting the ICL could modify the properties derived for the BGG, we also exploited the r band images processed to preserve all low surface brightness features (hereafter referred to as rLSB).
As explained e.g. by \cite{Zemaitis+23}, the images were obtained using an observing technique that is optimised for low surface brightness (LSB) surveys at CFHT \citep[e.g.,][]{Ferrarese+12,Duc+15}. 

There were 19 FGs available in rLSB.
For all these objects, we extracted images of 4000$\times$4000 pixels$^2$ centred on the BGG, either just by cropping the tiled UNIONS rLSB images, or by first assembling two to four tiled images with the SExtractor and SWarp softwares \citep[][]{BertinArnouts96,Bertin+02,Bertin06}\footnote{https://www.astromatic.net/}
and then cropping them to this size. In the latter case, special care was taken to assemble tiles that had been background subtracted in order not to lose faint signal in the outskirts of BGGs.

\subsection{Control sample of non-fossil groups} 

In order to compare the properties of BGGs of FGs with those of non-fossil groups (hereafter non-FGs), we built a catalogue of non-FGs in a similar way, but this time imposing a magnitude difference between the brightest and second brightest galaxies smaller than 2. This condition, together with that of $M_{halo}>10^{13}$~M$_\odot$ gave 100 non-FG candidates. None of them had a match in the XCLASS X-ray catalogue of N. Clerc.

The catalogues of the 87 FG candidates and of the 100 non-FGs are available in electronic form at the CDS via anonymous ftp to cdsarc.u-strasbg.fr (130.79.128.5)
or via http://cdsweb.u-strasbg.fr/cgi-bin/qcat?J/A+A/. They contain the following columns: number, RA, DEC, spectroscopic redshift, BGG stellar mass, absolute magnitude in the g and r SDSS bands, virial radius, and halo mass. 
Among the 100 non-FGs, we chose 30 non-FGs having good quality UNIONS images in u, r, and rLSB.
They are listed in Table~\ref{tab:30NonFG}.
This subsample of non-FGs is comparable in number to that of FG candidates for which UNIONS data are available. The BGGs of this control sample were chosen to match as well as possible the absolute magnitude and colour histograms of our FG sample. This can be checked by looking at Figs.~\ref{fig:histomasshalo88} - \ref{fig:histomags}. A Kolmogorov-Smirnov test on the FG and non-FG samples shows that these two samples have only a 40\% probability to be statistically different.

\begin{figure}[h]
\begin{center}
\includegraphics[width=0.49\textwidth]{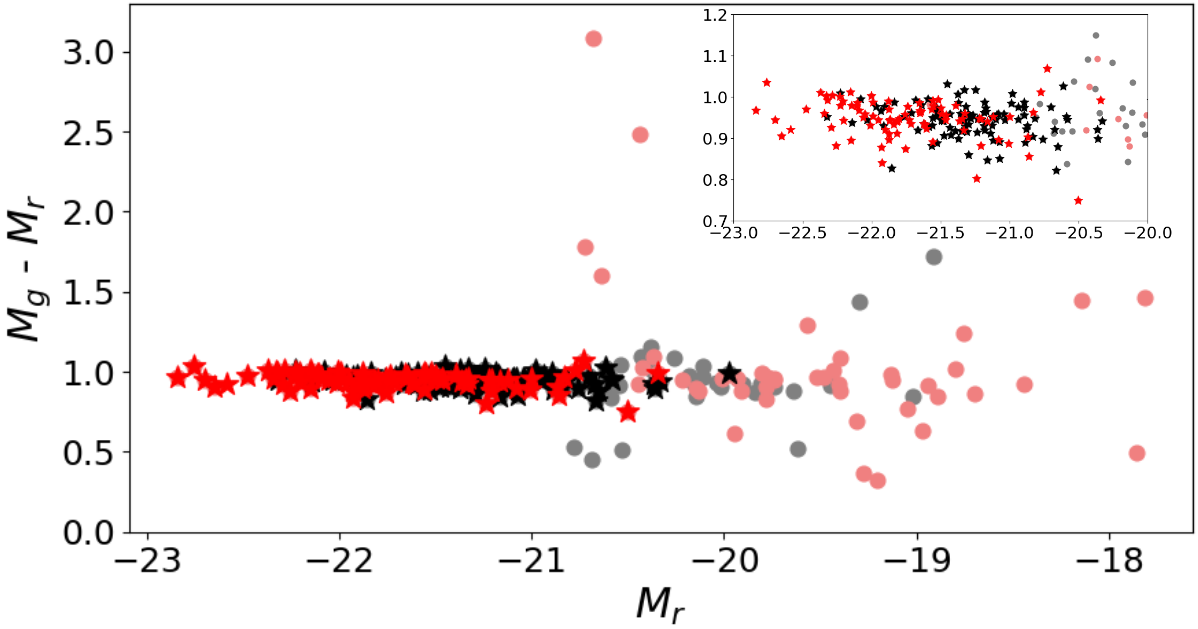}
\end{center}
\caption{(g-r) colour magnitude diagram for four samples: light red dots: galaxies within 1~arcmin of FG BGGs; red stars: FG BGGs; grey dots: all galaxies within 1~arcmin of non-FG BGGs; black stars: non-FG BGGs. The x-axis is the absolute magnitude in the r band. A zoomed-in version of the plot at the brightest magnitudes is shown on the top right.}
\label{fig:coulmag}
\end{figure}

For each FG and non-FG, we extracted from the Tinker catalogue (which gives absolute magnitudes in the g and r bands) all the galaxies belonging to the same halo within 1~arcmin around the BGG. The colour-magnitude diagram shown in Fig.~\ref{fig:coulmag} shows that all FG BGGs follow a very well-defined  sequence. With a colour g-r=$-1.29$, only one BGG is bluer than g-r=0 and falls outside the plot; it appears to be a spiral galaxy, but we kept it in the sample. BGGs of non-FGs surprisingly follow very well this sequence, whereas the galaxies which are not BGGs tend to be fainter and show a notably larger dispersion.

\begin{figure*}[h!]
\begin{center}
\includegraphics[width=1.0\textwidth]{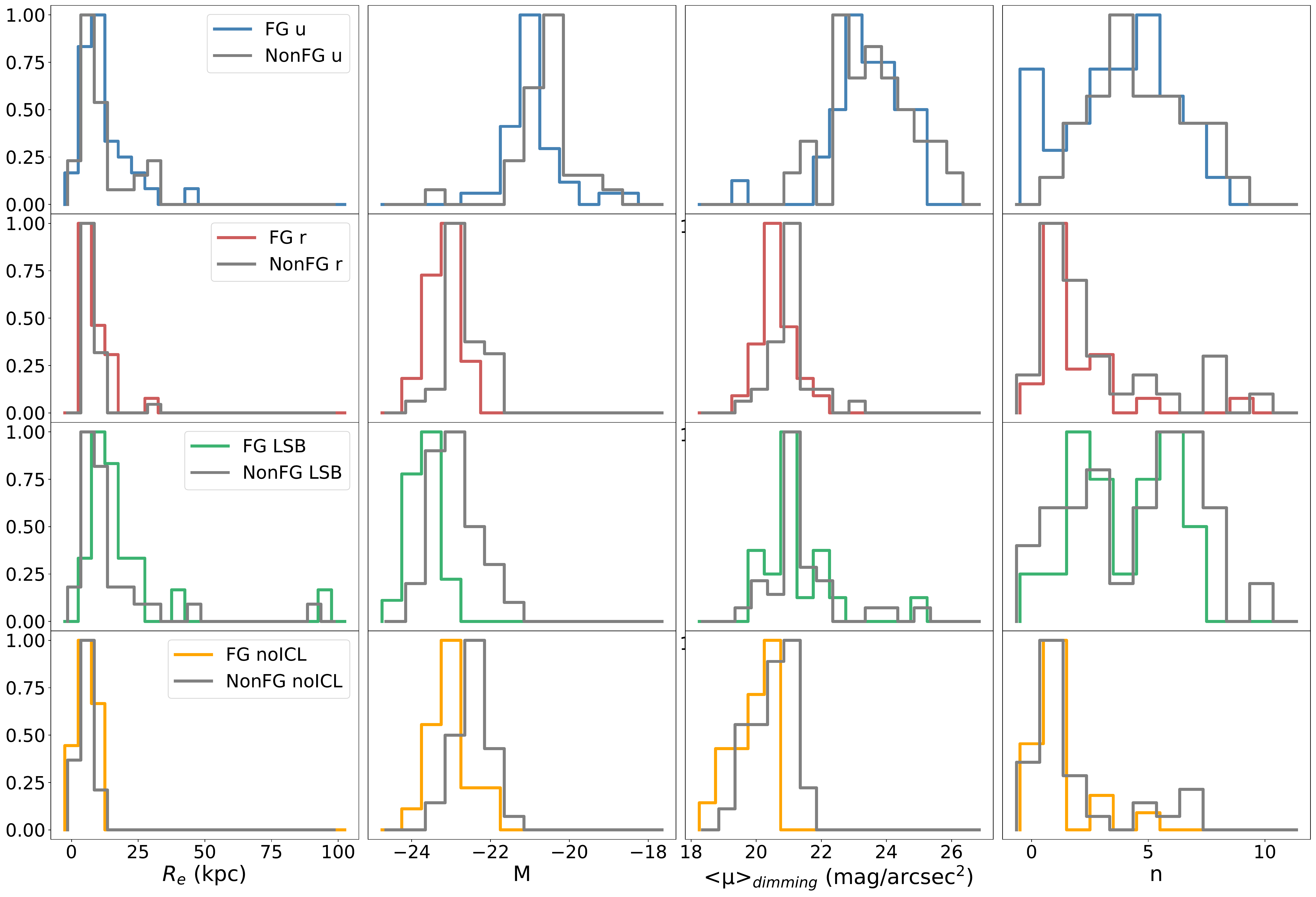}
\end{center}
\caption{From left to right: normalized histograms of the effective radius, absolute magnitude, mean surface brightness corrected for cosmological dimming, and S\'ersic index $n$. From top to bottom: 35 candidate FGs (in blue) and 30 non-FGs (grey) in the u band; 25 candidate FGs (in red) and 30 non-FGs (grey) in the r band; 19 candidate FGs (green) and 30 non-FGs (grey) in the rLSB band; same as above in the rLSB band after ICL subtraction (yellow for FGs).}
\label{fig:histoparam}
\end{figure*}

\section{Morphological properties of the FG BGGs}
\label{sec:morpho}

\subsection{Method} 

We now consider the 25 and 30 BGGs of FGs and non-FGs for which r band data are available in UNIONS.
Following \cite{Chu+21} and \cite{Chu+22}, the 2D profile of each BGG was fit with GALFIT \citep[][]{peng2002detailed}  with a single S\'ersic model and with a double S\'ersic model. GALFIT initial parameters were obtained with SExtractor, using a bulge + disk model. A mask and a point spread function were created following the method described in \cite{Chu+21}.
The choice between a single and double S\'ersic model was made based on the statistical F-test \citep[][]{margalef2016MNRAS.461.2728M} and the computed p-value, which indicates if complexifying the model, i.e increasing the number of degrees of freedom, is necessary. Since the current sample has the same resolution as \cite{Chu+22}, we adopt the same p-value limit to distinguish between the two different models: $P_0$ = 0.15.

\subsection{Morphological properties of FGs} 

\begin{table*}[h!]
\centering
\caption{Number of BGGs of fossil and non-fossil groups for which one or two S\'ersic laws are needed to fit their 2D profiles in the various bands: u, r, rLSB, and rLSB after subtraction of the ICL contribution.}
\label{tab:Sersic}
\begin{tabular}{r|rrrr|rrrr}
\hline \hline
& &&  FG & &  & & Non-FG & \\ 
\hline
 & u & r & rLSB & rLSB-ICL & u & r & rLSB & rLSB-ICL \\
\hline
One-S\'ersic &  29 &  2 &  9 & 3 & 27 &  6 & 14 & 6 \\
Two-S\'ersic &   6 & 23 & 10 & 16 & 3 & 24 & 16 & 24\\
\hline
\end{tabular}
\end{table*}

The numbers of BGGs of FGs and non-FGs for which one S\'ersic law (S1) or two S\'ersic laws (S2) are needed to fit their  2D profiles in the various bands are given in Table~\ref{tab:Sersic}.
For FGs, we can see that in the u band most BGGs can be fit with a single S\'ersic (83\%). In contrast, a major part of the profiles in the r band require two components (92\%).
If we now consider the BGGs from the rLSB data, only 53\% require two S\'ersics.

The fact that a single S\'ersic fits most of the u band images of BGGs can be due to the lower signal to noise ratio in the u band that makes the detection of a faint outer component difficult, while this outer low surface brightness component is better detected in the deeper r band images. Following the same interpretation, it appears surprising that the percentage of BGGs better fit by two S\'ersics is smaller in the deeper rLSB data than in the r band data. This can be explained by the fact that the Elixir-LSB pipeline, during the refined background
subtraction procedure, removes light from around the BGGs. We cannot guarantee whether this light has BGG origin, is due to instrumental contamination, or a mixture of both. This nevertheless means that part of our two S\'ersic fits in the r band are potentially contaminated by MegaCam scattered light. We decide to keep both bands for comparison purposes: the r band, with more flux but potentially contaminated, and the rLSB band that is 'cleaner' but might be missing light from the BGG outer profiles.

The same 2D profile fitting was applied to the control sample of non-FGs. 
As for FGs, we see in Table~\ref{tab:Sersic} that most BGGs in the u band are fit with a single S\'ersic (90\%) while in the r band 80\% are better fit with two S\'ersics, and in rLSB, 53\% are better fit with two S\'ersics. The difference between FGs and non-FGs is therefore of a few percent, a value that is not significant in view of the dispersion. 

As will be described in Sect.~\ref{sec:ICL}, we computed the ICL contribution and subtracted it from the BGGs to see how the S\'ersic parameters changed. If we look at Table~\ref{tab:Sersic}, we can see that only three FG BGGs out of 19 (16\%) can be fit with a single S\'ersic, and for non-FG BGGs six out of 30 BGGs (20\%) are fit with a single S\'ersic. The subtraction of the ICL contribution therefore does not make the second S\'ersic component disappear, as already observed by \cite{Chu+22}. The effect of ICL on the morphological properties of BGGs will be discussed further in Sect.~\ref{sec:ICL}.
Unexpectedly, we find more BGGs of FGs and non-FGs that are better modeled by two-S\'ersic profiles on ICL-subtracted rLSB images than on rLSB images. This will be discussed in Sect.~\ref{sec:discu}.

As seen in Fig.~\ref{fig:histoparam}, the effective radii are comparable for the BGGs of FGs and non-FGs in all filters.
The absolute magnitudes of FG BGGs appear brighter (by up to one magnitude) than those of non-FG BGGs. 
The same is observed for the mean surface brightnesses, except in the u filter, where they appear similar.

The histograms of the S\'ersic indices n are more concentrated towards low values in r for FGs: only 2 FGs out of 25 (8\%) have $n>$4 while 8 non-FGs out of 30 (27\%) have n$>$4. In rLSB, there are 10 FGs out of 19 (53\%) and 18 non-FGs out of 30 (60\%) with n$>$4. This high value of $n$ may be linked to the presence of ICL (see Sect.~\ref{sec:ICL}).

Because the difference in S\'ersic indices between FGs and non-FGs is not apparent in all bands, and the effective radii are about the same, the only clear difference that we find between the physical properties of FGs and non-FGs is that FG BGGs are intrinsically brighter. This may be due to a selection bias, as our sample of FG BGGs is overall brighter than our sample of non-FG BGGs (see Fig.~\ref{fig:histomags}). Despite our efforts to build two samples with similar properties, we could not find as many non-FG BGGs as bright as FG BGGs. This may indicate that, in a given group mass range, FG BGGs are indeed brighter and most of the mass of the group is concentrated in the BGG.

\subsection{Kormendy relation}

\begin{table*}[h!]
\caption{Slopes (a) with their error ($\sigma _a$) and intercepts (b) with their error ($\sigma _b$) for the Kormendy relation.
For each filter, we give the values without (u, r, rLSB, rLSB-ICL) and with (udimm, rdimm, rLSBdim, rLSB-ICLdimm) redshift dimming correction. Magnitudes were converted to the CFHTLS i magnitudes to compare the present results with those of \cite{Chu+22}.}
\centering
\label{tab:kormendy}
\begin{tabular}{l|rrrr|rrrr}
\hline\hline
      & &      FG & & & &                  Non-FG & &  \\
\hline
      &  a & $\sigma_a$ & b & $\sigma_b$ &   a & $\sigma_a$ & b & $\sigma_b$ \\
      \hline
u      & 3.41 & 0.28 & 17.15 & 0.31 & 3.72 & 0.44 & 17.13 & 0.47 \\
udimm  & 3.31 & 0.30 & 16.95 & 0.33 & 3.68 & 0.44 & 16.83 & 0.47 \\
r      & 3.80 & 0.32 & 16.82 & 0.34 & 3.95 & 0.68 & 17.24 & 0.69 \\
rdimm  & 3.26 & 0.39 & 17.07 & 0.41 & 3.76 & 0.68 & 17.09 & 0.69 \\
rLSB      & 4.48 & 0.23 & 15.79 & 0.29 & 4.03 & 0.31 & 17.07 & 0.35 \\
rLSBdimm  & 4.47 & 0.30 & 15.47 & 0.38 & 4.00 & 0.31 & 16.75 & 0.35 \\
rLSB-ICL      & 3.41 & 0.30 & 17.16 & 0.27 & 3.69 & 0.44 & 17.53 & 0.39 \\
rLSB-ICLdimm  & 3.10 & 0.36 & 17.10 & 0.32 & 3.44 & 0.44 & 17.41 & 0.38 \\
\hline
\end{tabular}
\end{table*}

We now consider the relation found by \cite{Kormendy77}
(mean effective brightness $<\mu>$ as a function of effective radius $R_e$), that can be fit with the following law:
$$<\mu>=(a\pm \sigma _a)\times log(R_e) + (b\pm \sigma _b).$$

The slopes and intercepts with their errors are given in Table~\ref{tab:kormendy} in the u, r, rLSB and ICL subtracted rLSB bands, before and after correction for cosmological dimming, applied as in \cite{Chu+22}. For better comparison with that study, we convert the surface brightnesses measured in the u, r, rLSB and ICL subtracted rLSB bands into those in the i band using the \cite{Fukugita+95} tables.
We can see in Table~\ref{tab:kormendy} that the dispersion is smaller for FGs than for non-FGs in all bands, and that the relations in u and r are parallel within the dispersion. Except in the rLSB band, the slopes are larger for non-FGs than for FGs, but in view of the dispersion this difference is not significant.

\begin{figure}[h]
\includegraphics[width=9.cm,angle=0]{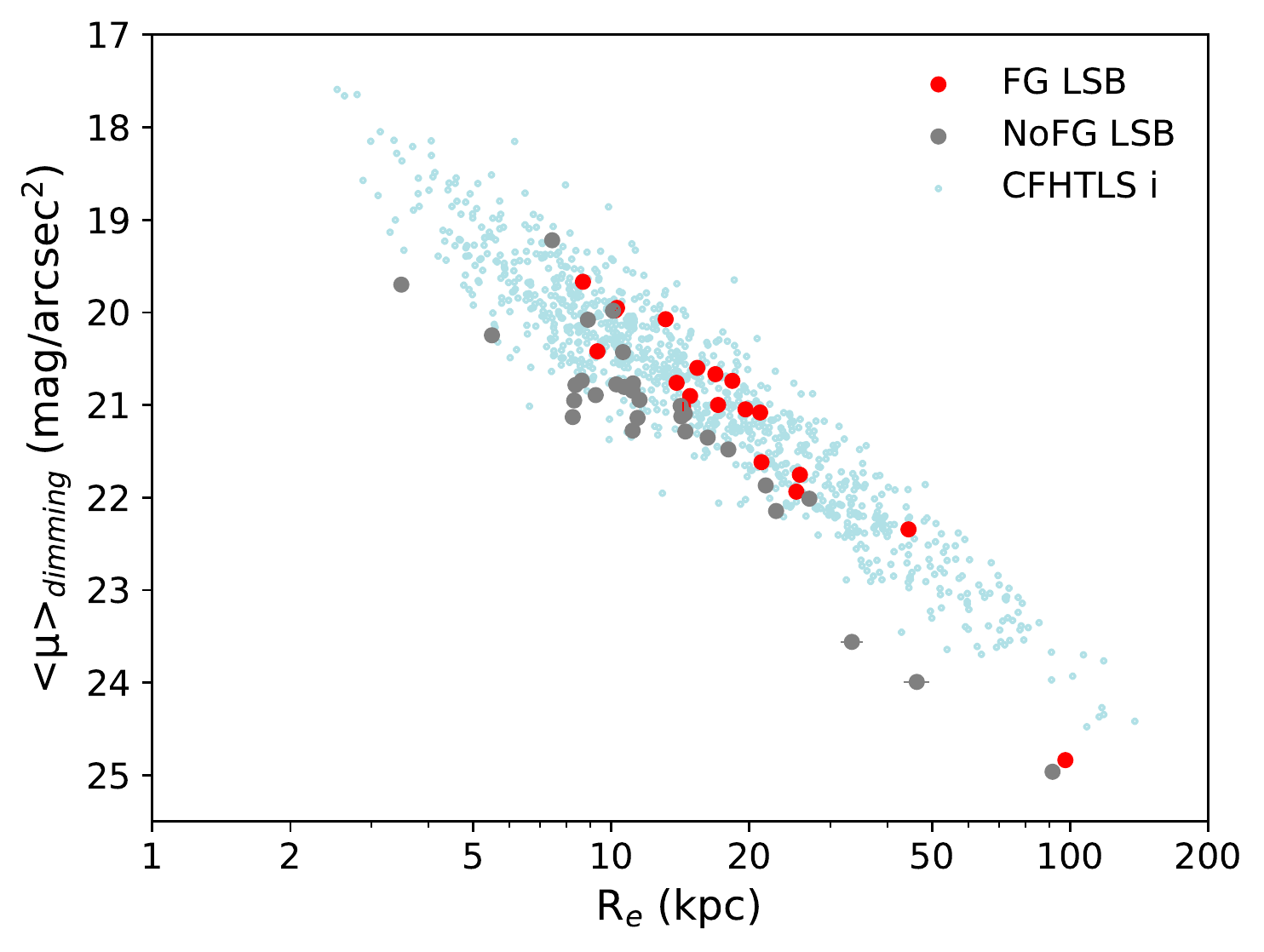}
\includegraphics[width=9.cm,angle=0]{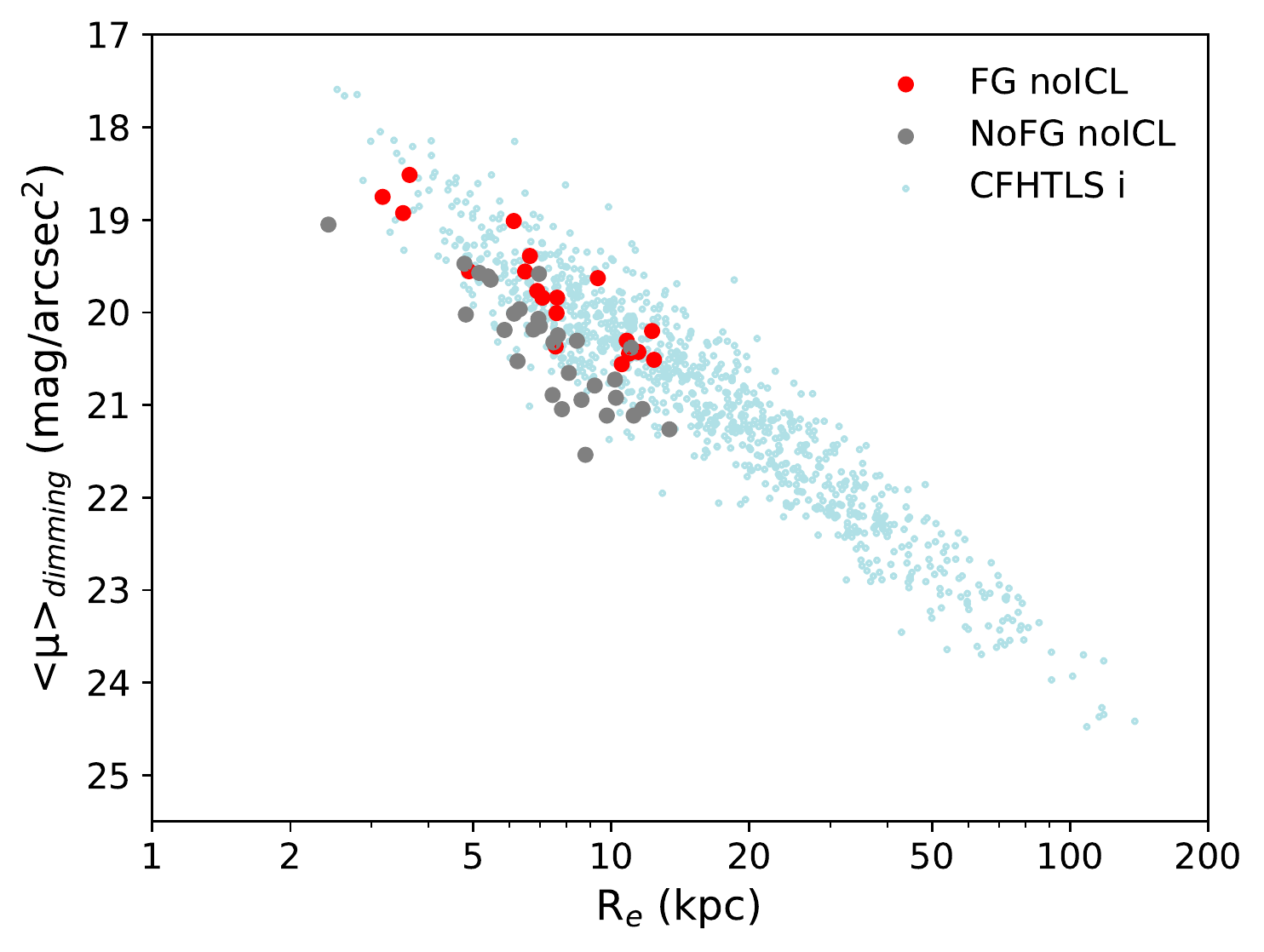}
\caption{Kormendy relation for FGs in red and Non-FGs in grey, superimposed on the relation found for almost one thousand BCGs by \cite{Chu+22} in cyan. Both plots are obtained in the rLSB band, the top plot without ICL subtraction and the bottom plot after ICL subtraction.
The points for FGs and non-FGs have been shifted from the r to the i band, and all points are corrected for cosmological dimming.}
  \label{fig:kormendy}
\end{figure}

The Kormendy relation derived in the i band for almost one thousand BCGs by \cite{Chu+22} was:
$<\mu >= (3.34 \pm 0.05) \times log(R_e) + (18.65 \pm 0.07)$ and
$<\mu >= (3.49 \pm 0.04)\times log(R_e) + (16.72 \pm 0.05)$ 
before and after correction for cosmological dimming respectively. This relation was found to be independent of the model used (one or two S\'ersic profiles).
If we compare the Kormendy relations found here for FGs and non-FGs to those of \cite{Chu+22}, we see that FGs are located on this relation, while non-FGs are located in majority below the relation, as illustrated in Fig.~\ref{fig:kormendy}, and this is true both before and after ICL subtraction. However, in view of the large dispersion the difference in the intercepts is not statistically significant. 

Therefore the BGGs of FGs appear to have properties closer to those of BCGs than non-FG BGGs, suggesting that BGGs in FGs and BCGs have  undergone comparable evolutions, while non-FGs have evolved somewhat differently, reaching fainter surface brightnesses.

\begin{figure*}
\begin{center}
\includegraphics[width=0.8\textwidth]{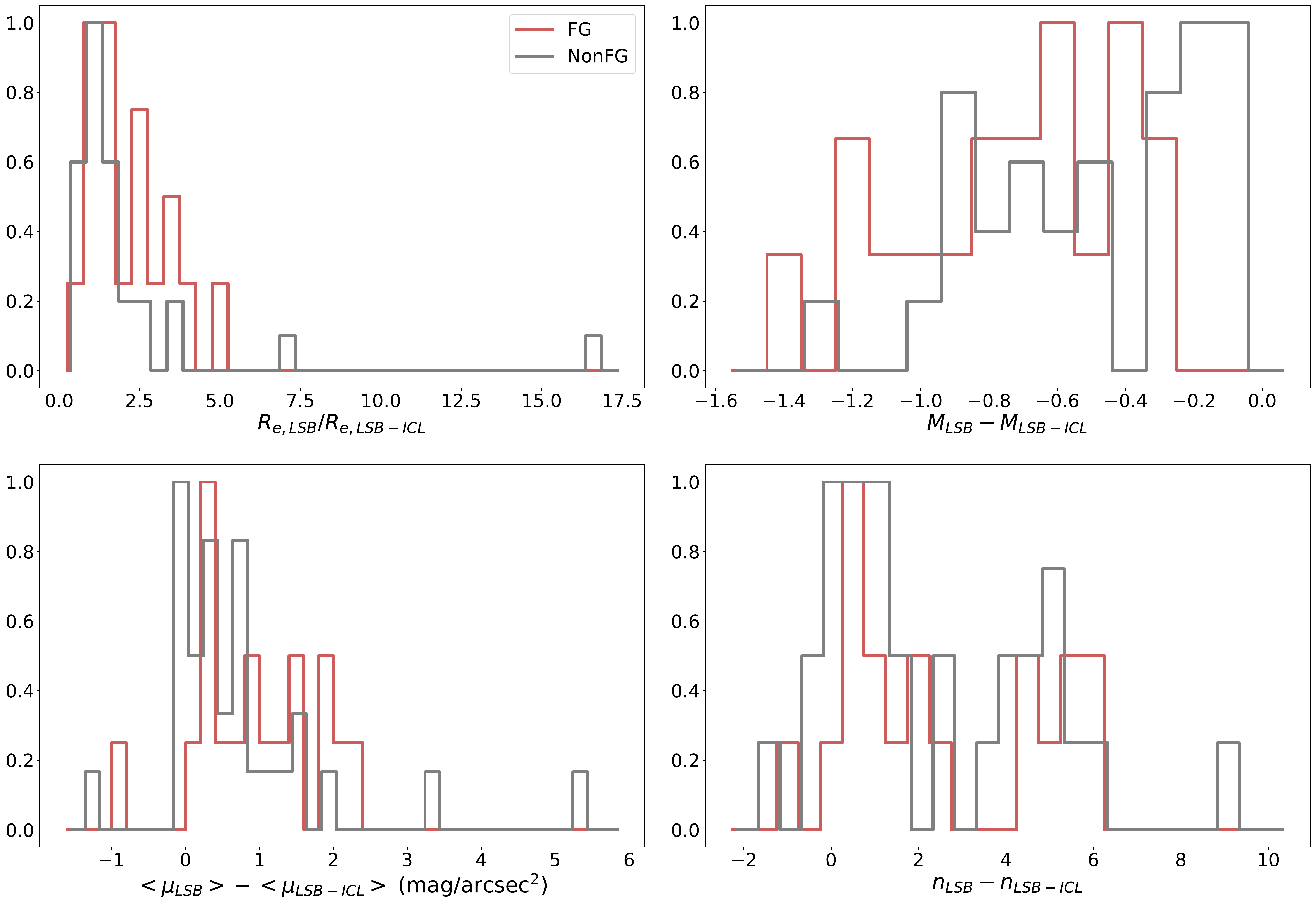}
\end{center}
\caption{Normalized histograms of the difference between the values obtained before and after subtraction of the ICL for the 19 candidate FGs (in red) and 30 non-FGs (in grey): effective radius (top left), absolute magnitude (top right), mean surface brightness (bottom left), and S\'ersic index $n$ (bottom right).}
\label{fig:histoLSB-noICL}
\end{figure*}

\subsection{Influence of the ICL contribution on the estimation of the morphological properties of FGs}
\label{sec:ICL}

\begin{table*}
\centering
\centering
\caption{Mean morphological parameters (with their dispersion in parentheses) for FGs and non-FGs in the u, r, and rLSB bands, and after subtracting the ICL contribution. The columns are: effective radius R$_e$, absolute magnitude M$_{abs}$, mean surface brightness $\mu _{mean}$ corrected for cosmological dimming, and S\'ersic parameter $n$.}
\label{tab:ICLnoICL}
\begin{tabular}{lrrrr}
\hline \hline
  &  R$_e$~~~ & M$_{abs}$~~ &  $\mu _{mean}$~~  &  $n$~~~~ \\ 
\hline
FG u & 14.0 (8.8) & -20.6 (0.8) & 23.7 (1.0) & 4.2 (2.3) \\
Non-FG u & 13.2 (8.7) & -20.4 (0.7) & 23.8 (1.2) & 5.2 (2.0) \\
FG r & 11.7 (4.8) & -23.0 (0.4) & 20.9 (0.5) & 2.3 (1.9) \\
Non-FG r & 10.6 (4.6) & -22.4 (0.5) & 21.3 (0.6) & 3.3 (2.5) \\
FG rLSB & 22.0 (19.5) & -23.4 (0.4) & 21.5 (1.1) & 4.4 (2.1) \\
Non-FG rLSB & 16.7 (16.4) & -22.6 (0.5) & 21.6 (1.2) & 5.0 (2.7) \\
FG no ICL & 7.9 (2.9) & -22.8 (0.5) & 20.2 (0.6) & 1.6 (1.2) \\
Non-FG no ICL & 7.7 (2.4) & -22.2 (0.4) & 20.8 (0.6) & 2.4 (2.1) \\
\hline
\end{tabular}
\end{table*}

\cite{Chu+22} have estimated the importance of the ICL contribution on the morphological properties of seven BCGs computed with GALFIT. For this, they fit the 2D properties of the BCGs on the original images, and then subtracted to these images the contribution of the ICL derived by \cite{Jimenez+18} and fit again the BCGs. Their main result was that two S\'ersics were still necessary to fit the BCGs, and therefore that the need for a second component could not only be attributable to ICL. In all cases, the effective radii increase in the presence of ICL, some of them drastically (by an order of magnitude).
For all seven BCGs, the absolute magnitude of the external component is fainter after removing the ICL (with a difference that can reach two magnitudes). After subtracting ICL, BCGs also have
brighter effective surface brightnesses, with a difference that can almost reach 3~mag~arcsec$^{-2}$. 

Similarly, to better quantify the effect of the ICL in the present study, we now compare the values of the various physical parameters galaxy by galaxy before and after ICL subtraction. This demands an estimation of the ICL contribution for each FG. For this, we use \texttt{DAWIS} \citep[Detection Algorithm with Wavelets for Intracluster light Studies; ][]{Ellien+21}, a wavelet based algorithm optimized for the detection and characterization of LSB structures in deep optical images. In a nutshell, \texttt{DAWIS} applies a wavelet transform to the analysed image and detects sources in the new wavelet representation. The light profiles in the original image of these sources are then modeled from their wavelet representation through a restoration procedure. \texttt{DAWIS} follows a semi-greedy procedure, meaning that sources are restored iteratively. The brightest detected sources are modeled first and removed from the image, and the whole wavelet procedure is then applied again on the residual image. New sources are detected, modeled and removed, until convergence to a noise only residual map. This provides a refined modeling of all detected sources, from very bright objects down to very faint structures. Note that the detected sources are not necessarily entire astrophysical objects, but rather sub-structures of such objects. \texttt{DAWIS} was applied to each FG for which we have an image in the rLSB band, providing for each of them a list of sources and their light profiles.

For each BGG, we extracted from the \texttt{DAWIS} output list a sub-list of sources that are identified as contributing to ICL. This sub-list was estimated using the following criteria: the source must be unimodal and centered on the BGG center, and its spatial radial extent must be greater or equal to a given radius. We tried different values for the spatial extent, going from 40~kpc up to 100~kpc. Low values lead to brighter ICL contribution, contaminated by smaller sources such as pieces of foreground and background galaxies. For our study we chose to remove a clean ICL contribution, potentially missing some of it, rather than a contribution contaminated by galaxies, and set the radius to 100~kpc. The light profiles of these sources are summed into a 2D ICL profile, and this profile is then subtracted from the BGG light profile. More details about this procedure and \texttt{DAWIS} can be found in \citet{Ellien+21} and Brough et al. (in preparation).

The results after ICL subtraction are illustrated in Fig.~\ref{fig:histoLSB-noICL} and Table~\ref{tab:ICLnoICL}, We can see that for FGs, the ICL increases the effective radius by a mean factor of 3.7, and the ICL adds light to the BGG by up to 1.3 magnitude. With ICL, the mean surface brightness can be up to 5~mag~arcsec$^{-2}$ brighter, with a mean at 1.3, and the S\'ersic index $n$ tends to be larger (by 2.8 on average). There seems to be a bimodality in the S\'ersic index histogram. The peak at $n<2$ comes from BGGs that were better modeled with two S\'ersic profiles before and after ICL subtraction, and the peak at $n>3$ originates from 1-S\'ersic profile BGGs on LSB images that become 2-S\'ersic BGGs after ICL subtraction. 

For non-FGs, the ICL increases the effective radius by a mean factor of 2.4 and the absolute magnitude with ICL is up to 1.2 magnitude brighter. The mean surface brightness before subtraction of the ICL can be up to 1.2 mag~arcsec$^{-2}$ fainter, with a mean of about 0.8, and the S\'ersic index $n$ tends to be larger with ICL, with a large dispersion from one BGG to another.
We can note that when the ICL is subtracted, the S\'ersic index distributions become comparable for FGs and non-FGs (refer to Fig.~\ref{fig:histoparam}). 
The distributions of the properties of BGGs obtained for FGs and non-FGs after ICL subtraction also become more comparable to those measured in the r filter (see Fig.~\ref{fig:histoparam}). This is not surprising, as the r images before LSB processing should resemble rLSB images after the removal of low brightness features. However, we can note that there is still a significant difference 
between the properties measured on the r and rLSB-ICL images. Indeed, BGGs on the r filter appear bigger, brighter, and have fainter mean surface brightnesses and higher S\'ersic indices (see Table~\ref{tab:ICLnoICL}) than on rLSB-ICL images, all which indicate the presence of ICL even on the r images.
This illustrates the fact that ICL modifies somewhat the apparent morphological properties, and that its contribution should be carefully subtracted before analysing the physical properties of BCGs and BGGs.

\section{Stellar populations of BGGs}
\label{sec:popstell}

\subsection{Data}

We retrieved the spectra of the FG and non-FG BGGs of our sample in the SDSS. Among the 87 FGs, nine spectra were not available, leaving us with 78 BGGs. Among the 100 non-FGs, 4 spectra were not available, so we analysed 96 spectra.
We fit these spectra with Firefly \citep{Wilkinson2017MNRAS.472.4297W} and eliminated the spectra corresponding to AGN as classified by the SDSS: 13 in FGs and 14 in non-FGs. This was done because it is difficult to extract stellar populations from spectra dominated by an AGN.
The AGN fractions are very similar between BGGs of FGs and non-FGs, 20\% and 17\% respectively.
We were therefore left with 66 FG BGG and 82 non-FG BGG spectra. Since \cite{Wilkinson2017MNRAS.472.4297W} have shown that Kroupa and Salpeter IMFs gave comparable results, we limited our analysis to a Kroupa IMF. We used the STELIB and MILES stellar libraries, to estimate the robustness of our results. 

\subsection{Results}
\label{sec:results}

\begin{figure}[h]
\begin{center}
\includegraphics[width=0.4\textwidth]{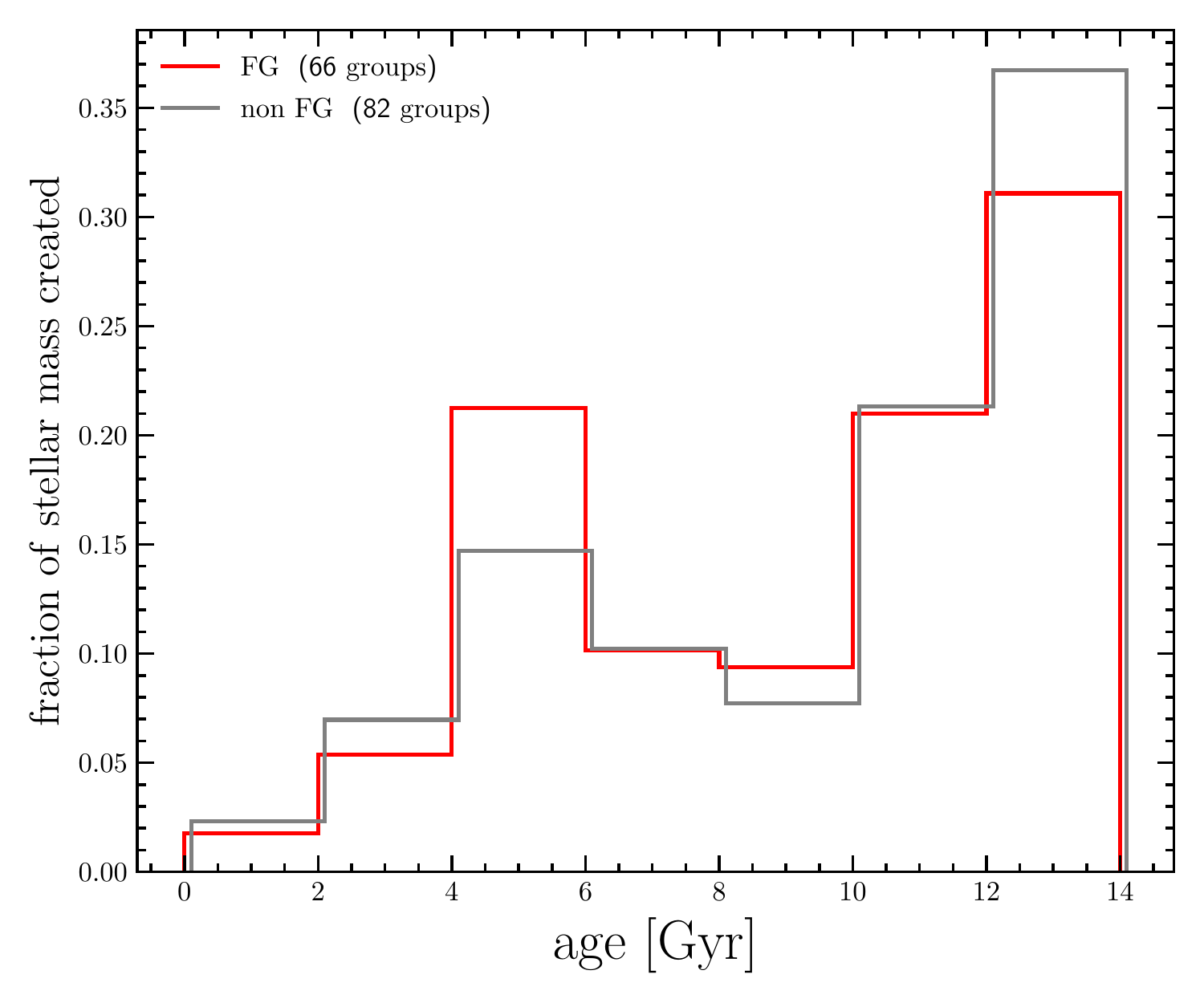}
\includegraphics[width=0.4\textwidth]{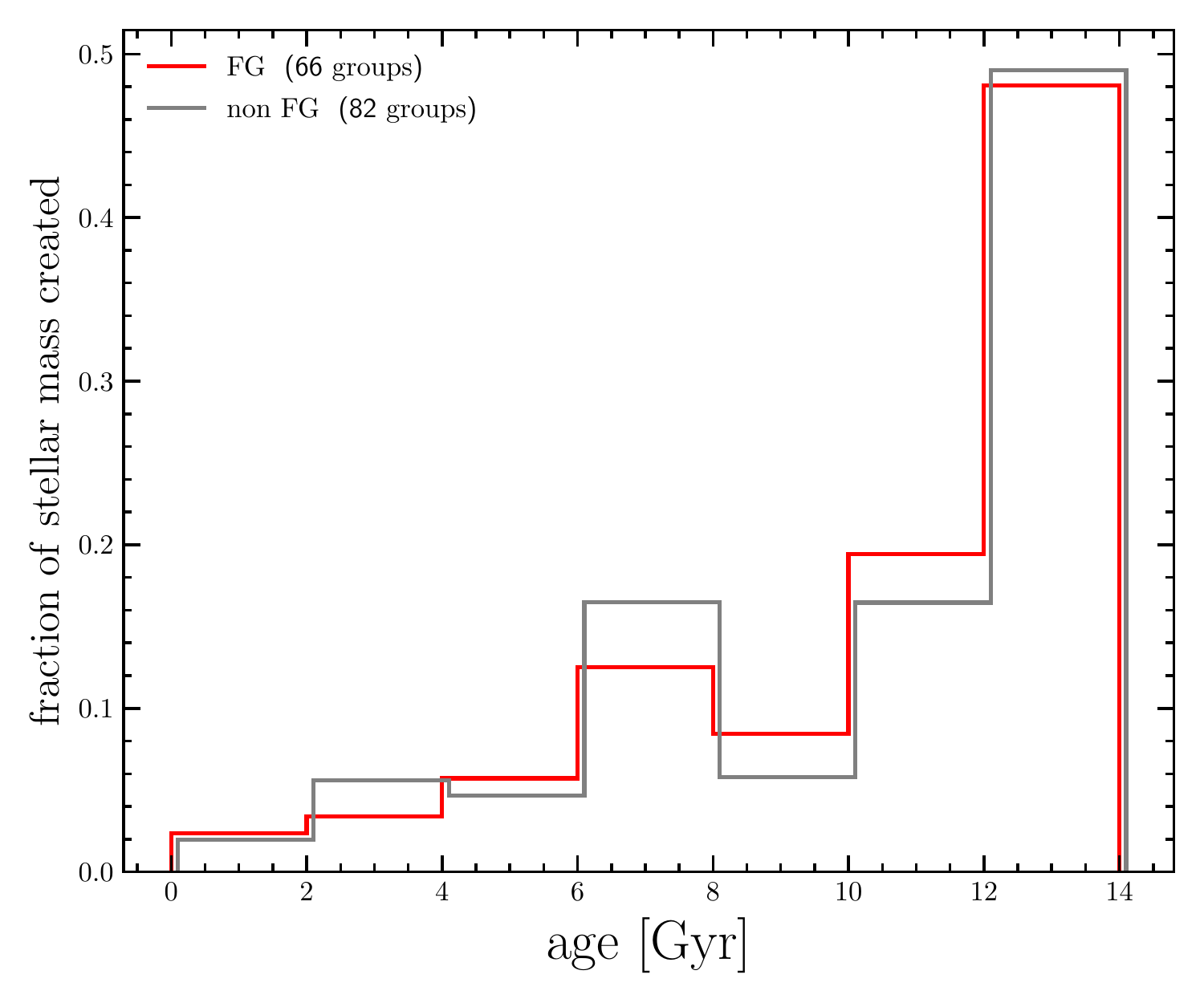}
\end{center}
\caption{Histograms of the fraction of stellar mass created as a function of age for FGs (red) and non-FGs (grey), for a Kroupa IMF, using the MILES (top) and STELIB (bottom) libraries.}
\label{fig:histoages}
\end{figure}

We find that the stellar populations of FG and non-FG BGGs cannot be distinguished with this relatively straightforward analysis. As an illustration, we show the fraction of stellar mass created as a function of age in Fig.~\ref{fig:histoages}.
Firefly fits the galaxy observed spectra using multiple star-formation bursts with various luminosities, masses, metallicities and ages. The software returns the number of bursts, and for each burst: its age, metallicity, and the fraction it contributed to the luminosity and stellar-mass of the galaxy. From these, we computed the fraction of stellar mass formed in a range of galaxy age by summing the fractions of total stellar-mass of all the star-formation bursts that occurred in the considered age range.

We find that 76\% and 71\% of the total stellar mass of FG and non-FG BGGs respectively was already formed 8~Gyr ago, using the STELIB library. These percentages become 61\% and 66\% for the MILES library.
These results agree with Fig.~\ref{fig:coulmag}, where we see that the BGGs of FGs and non-FGs have very similar colours. Therefore, although the morphological properties of FG and non-FG BGGs somewhat differ, this is not the case for their stellar populations.
Fig.~\ref{fig:histoages} shows that the choice between the MILES and STELIB libraries does not change much the results.

There are 39 BGGs in FGs with emission lines (among which 9 may have a broad H$\alpha$ component) out of 78. Each time H$\alpha$ is detected, it is associated
with the [NII]6584 emission line. For the BGGs of non-FGs, 30 out of 96 show emission lines (with no broad H$\alpha$ line apparent). 

These emission lines can originate from star-formation processes (e.g. driven by
recent merging events) or from AGN activity (even if no broad H$\alpha$ line is detected). However, the spectroscopy available to us comes from the SDSS survey, with a fiber of 3~arcsec diameter. Given the angular extent of the considered BGGs, at most 10$\%$ of their area within a Petrosian radius is covered by the fiber, and mainly in the center of the galaxies. Detected emission lines are therefore probably coming from AGN activity, with rare cases 
where it can be produced by star-forming processes detected in the disk. 

The respective percentages of H$\alpha$-[NII] emission line galaxies are 50\% for FGs and 31\% for non-FGs. This suggests that BGGs contain more frequently an AGN in the centers of FGs than in non-FGs.

\subsection{Emission lines in FG BGGs: the case of NGC~4104}

As seen in the previous section, a non negligible part of our FG BGGs show emission lines and we think that these are mainly due to AGN activity. In order to investigate more precisely the location of the emitting regions in BGGs and see if some merging-induced star-formation activity can be detected, we concentrated on a specific example: NGC~4104 (not a member of our present sample of FG candidates). 

\begin{figure}[h]
\includegraphics[width=9.cm,angle=0]{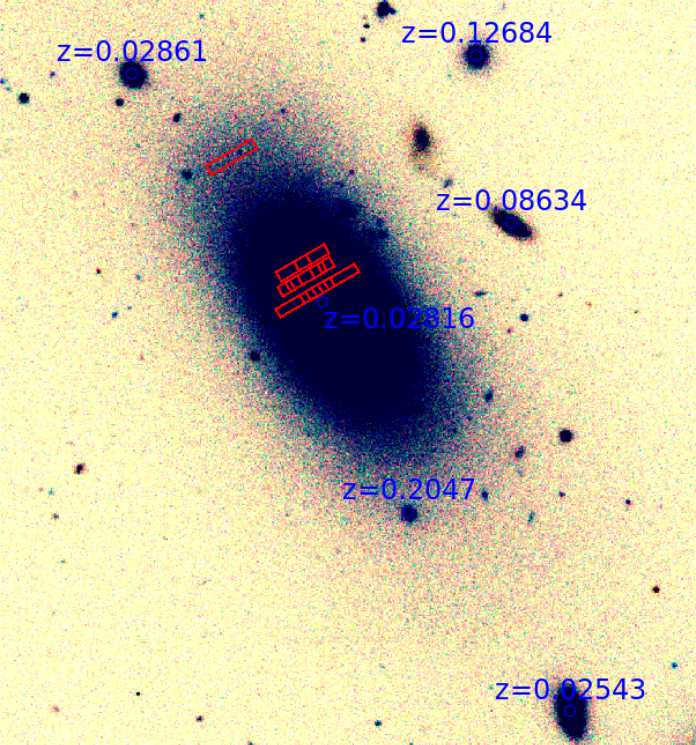}
\caption{SDSS g', r', i' trichromic image
  ($\sim$2.9'$\times$3.5') of NGC~4104. 
  The redshifts indicated come from the SDSS.
  The red rectangles show the regions where we extracted MISTRAL spectra. }
  \label{fig:image_N4104}
\end{figure}

\begin{figure}[h]
\includegraphics[width=9.cm,angle=0]{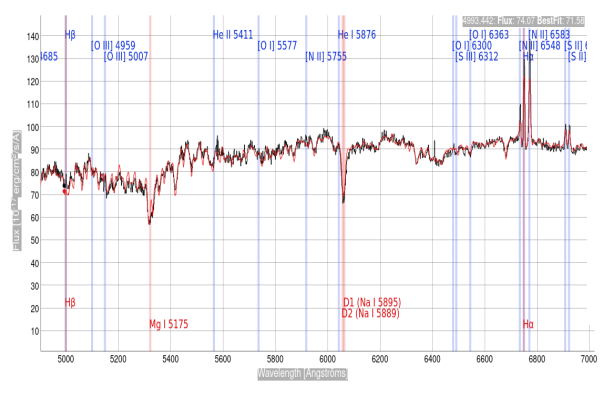}
\caption{SDSS spectroscopy of NGC~4104, with emission
  (blue labels) and absorption (red labels) lines.}
  \label{fig:spectre}
\end{figure}

It is a well known galaxy (see Fig.~\ref{fig:image_N4104}) at z$\sim$0.02816, BGG of the eponym fossil group recently studied by e.g. \cite{LimaNeto+20}.
It constitutes a good laboratory to investigate the possibility of having merging-induced star-formation activity because it was most probably subject to relatively recent merging activity \citep[][]{LimaNeto+20}. Given the fact that it is a nearby galaxy, it is also easy to study in detail its spectral and morphological characteristics.

Despite the BGG status of NGC~4104 \citep[classified as S0 by the RC3 catalog][and as Elliptical by the
SDSS]{deVaucouleurs+91}, SDSS spectroscopy (identified as spec-2227-53820-0518 in the
Science Archive Server) clearly shows strong emission lines in its center (see Fig.~\ref{fig:image_N4104}
for the location of the SDSS spectroscopic extraction area). In this
region, we have a noticeable lack of emission at the H$\beta$ and
[OIII] wavelengths. The H$\alpha$, [NII] and [SII] lines are very
prominent, in addition to MgI and NaD absorption lines (see Fig.~\ref{fig:spectre}).

In order to investigate the origin of these emission lines, and to see
if they are present over the entire galaxy, we partially mapped
NGC~4104 at the Observatoire de Haute Provence with the MISTRAL
single-slit spectro-imager (see
http://www.obs-hp.fr/guide/mistral/MISTRAL\_spectrograph\_camera.shtml)
in March 2022. We used the blue setting (1 hour exposure, resolution
R$\sim$750, and covered the wavelength range [4200, 8000]\AA) to obtain
spectra in four different slits (see Fig.~\ref{fig:image_N4104}). The three slits close
to the galaxy center provided high enough signal-to-noise to allow
extraction of spectra in 15 different regions (see Fig.~\ref{fig:spectres_zoom}). 
The outermost slit has too low a signal-to-noise to detect any significant
lines (including emission lines).

We do not detect any emission line in the galaxy external
regions with the other slits. Only central regions (l1, l2, l4, c1,
c2, c3, c5, and c6 on Fig.~\ref{fig:spectres_zoom}) show [NII], [SII], and sometimes weak
to bright H$\alpha$ emission, very similarly to what is visible in the
SDSS spectrum. We show in Fig.~\ref{fig:spectres_zoom} a zoom on the [6700, 6940]\AA\ 
wavelength interval. The table GalSpecInfo of the SDSS DR17 database
lists NGC~4104 as an AGN, with an old stellar population of the
order of 13~Gyr.  In order to investigate more precisely the nature
of this galaxy we first applied the pipes$\_$vis visualization tool \citep{Leung+21}, based on the BAGPIPES tool \citep{Carnall+18} to the normalized SDSS spectrum. We selected a \cite{Wild+20} model, adding dust \citep{Charlot+00} and a nebular
component \citep{Leung+21}.

First, based on \cite{LimaNeto+20}, we assumed a stellar mass for
NGC~4104 of 10$^{11.3}$ M$_\odot$ and a redshift of 0.03. We also fixed
to 100 the $\alpha$ and $\beta$ slopes of the \cite{Wild+20} burst
(decline and incline steepness of the burst, index of double power
law), to 2.0 the $\eta_{dust}$ (additional scaling factor for dust in
star forming clouds), to 0.7 the n$_{CF00}$ (slope of the attenuation law) of
the dust component, and to 0.01~Gyr the $t_{bc}$ parameter in
pipes$\_$vis (time during which the star forming clouds remain).

We then fixed the galaxy velocity dispersion to 150~km/s in order to
reproduce the [SII] doublet aspect.
The log(U) value (ionization parameter) has to be lower than $-3.5$ in
order to have [NII] stronger than H$\alpha$. We fixed log(U) to
this value.

We had to introduce a passive population older than $\sim$6~Gyr to
explain the depth of the absorption lines. We fixed the age of the
Universe when the older population formed to 8~Gyr. We also fixed to
1~Gyr the SFR decay timescale of the older population.

To explain the absence of [OIII] emission and the weakness of
H$\beta$, we fixed the metallicity to Z=2.5 and the Av (extinction in V
band) to 2.

Finally, we added a small burst in the \citet{Wild+20} model, with 5$\%$
of the mass of the older population contained in the burst, $\sim$1~Gyr ago 
(12.8~Gyr after the Big Bang), in order to explain visible emission lines and to 
have deep enough MgI and NaD lines. Values contained in the burst of more than 
10$\%$ of the mass of the older population do not 
allow to reproduce the spectrum characteristics, whatever the values of the other parameters.

\begin{figure}[h]
\includegraphics[width=9.cm,angle=0]{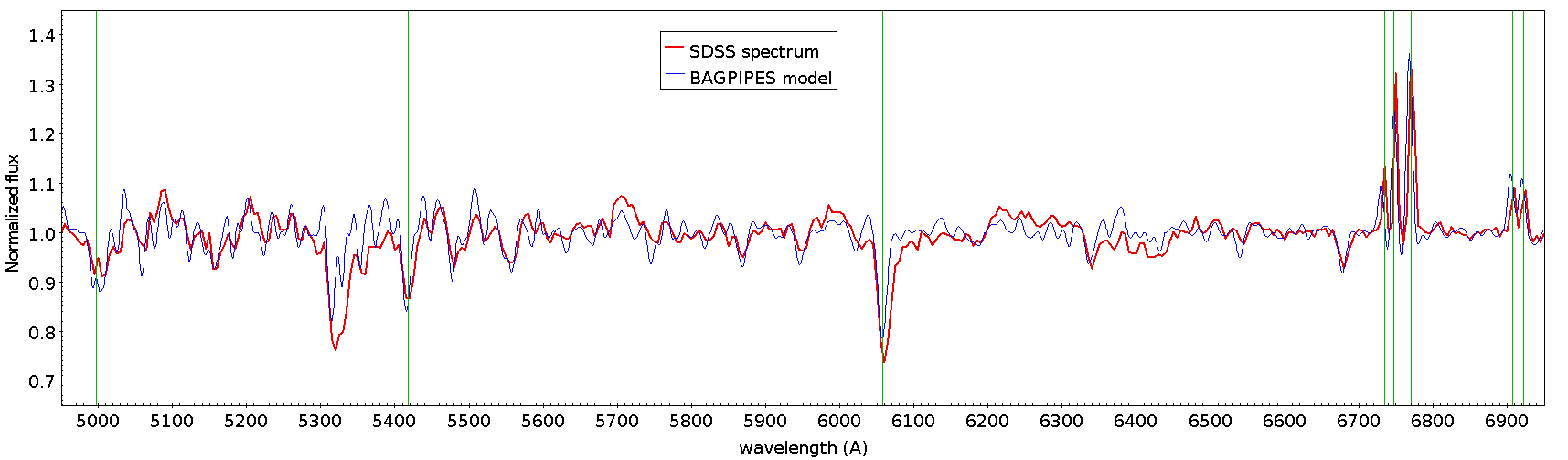}
\caption{A modelisation of the normalized SDSS
  spectrum of NGC~4104 with the pipes$\_$vis tool.}
  \label{fig:spectre_model}
\end{figure}

With this set of parameters, we are able to reasonably reproduce the
SDSS spectrum (see Fig.~\ref{fig:spectre_model}). This probably shows that we are dealing
with a relatively old stellar population ($\sim$6~Gyr old). This is
consistent with the 4-6~Gyr old merger proposed by \cite{LimaNeto+20}. Any emission lines induced by the old merger have probably now vanished.  In addition, a recent burst/merger ($\sim$~1~Gyr old) is also likely. It may possibly have reactivated the central AGN and be at
the origin of the emission lines presently observed in the centre of NGC~4104. The star-forming H$\alpha$ emission induced by this late merging event would require a few hundred Myr to also vanish, and this is not in contradiction with the age of the recent burst ($\sim$~1~Gyr).

As a comparison to the other FGs, we also applied Firefly to the SDSS spectrum of NGC~4104 with a Kroupa IMF and the STELIB and MILES stellar libraries. For these two models, we find that 73\% and 69\% of the total stellar population was formed more than 8~Gyr ago, which is consistent with the previous results. The subtraction of the stellar population fitting results in a pure-emission spectrum, that can be used to locate the line ratios in the BPT  \citep[]{BPT81} diagrams
\citep[][]{Kewley+13}. The line ratios [OIII]/H$\beta$ and [NII]/H$\alpha$ locate the nuclear spectrum of NGC~4104 in the region occupied by LINER nuclei.

\begin{figure}[h]
\includegraphics[width=9.cm,angle=0]{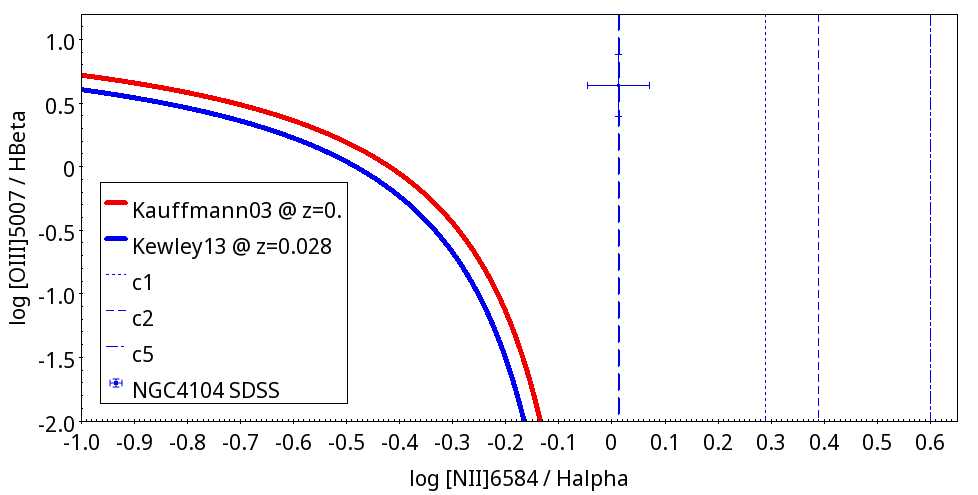}
\caption{log([OIII]5007/H$\beta$) versus
  log([NII]6584A/H$\alpha$) for the SDSS spectrum (central area, see
  e.g. Fig.~\ref{fig:spectres_zoom}) and the c1, c2, and c5 MISTRAL regions. Regions under the red and blue curves are normal galaxies, according to \cite{Kauffmann+03} and \cite{Kewley+13} respectively, 
  while regions above these curves are active objects (AGN, LINERS).}
  \label{fig:rapp-raies}
\end{figure}

We also investigated the AGN nature of the spectra in MISTRAL covered
regions where [NII] and H$\alpha$ emissions were strong (namely,
regions c1, c2 and c5 of Fig.~\ref{fig:spectres_zoom}). 
[OIII] and H$\beta$ were not
detectable, so Fig.~\ref{fig:rapp-raies} only shows vertical lines for these three
regions. We clearly see that, even without [OIII] and H$\beta$ lines, the
NGC~4104 c1, c2, c5, and central regions all have AGN characteristics.

As a conclusion, the emission lines of NGC~4104 originate mainly from the galaxy centre and each of the regions exhibiting emission lines have AGN characteristics. The remaining areas of the galaxy have passive
characteristics. The emission lines of NGC~4104 are therefore largely due to the AGN activity, in agreement with an AGN-origin for the emission lines that we detected in at least part of our general sample of BGGs (see previous section).

\section{Discussion and Conclusions}
\label{sec:discu}

The aim of this paper is to increase the number of known FGs to shed light on their formation process. Here, we increase the sample by 87 FG candidates confirmed spectroscopically, and with a high probability of being real FGs in view of their estimated large halo mass, though the confirmation of the X-ray condition on their X-ray luminosity
($\rm {L_X}>10^{42}$~erg~s$^{-1}$)
is still missing. For the FGs with UNIONS data available (35 in the u band, 25 in the r band and 19 in the rLSB band (r band treated with the Elixir-LSB software), we analyse the morphological properties of their brightest group galaxy (BGG), and compare these properties with those of a control sample of 30 non-FG BGGs. 

The 2D photometric fits of the BGGs made with GALFIT with one or two S\'ersic components show that a single S\'ersic component is sufficient in most objects in the u band, while two S\'ersics are needed in the r band, both for FGs and non-FGs. However, non-FGs cover a larger range of S\'ersic index $n$ than FGs. Therefore, FG BGGs appear to have homogeneous profiles whereas non-FG BGGs cover a larger range of profiles.

After the subtraction of ICL in the rLSB images, the fraction of two-S\'ersic BGGs increases. This is rather unexpected, since ICL forms a faint halo around the BGG and adds flux to the envelope of the BGG.
We could thus expect a higher fraction of two-S\'ersic BGGs when ICL is included. However, the ICL distribution does not necessarily follow a S\'ersic profile.
In a more general way, it seems that S\'ersic profiles generally do not trace physical components, and only act as an analytical description of galaxies. 

FG BGGs appear to follow the Kormendy relation previously derived for BCGs by \cite{Chu+22}, whereas the relation for non-FG BGGs is shifted to fainter surface brightnesses. This implies that FG BGGs have morphological properties closer to those of BCGs, and may have evolved similarly to BCGs.  \cite{Chu+21} have confirmed that BCGs have not evolved since z=1.8. Therefore, a similar evolution of FG BGGs and of BCGs favours the scenario in which FGs have formed long ago and have stopped evolving because their environment lacked galaxies that the group could accrete. 

We have analysed the properties of the stellar populations of FG and non-FG BGGs derived from the analysis of their SDSS spectra with Firefly, assuming a Kroupa IMF and using the STELIB and MILES stellar libraries. We find no significant difference between FGs and non-FGs (65 and 82 galaxies respectively), so the stellar populations of BGGs must have evolved in comparable ways in FGs and non-FGs.

Detailed observations of the FG NGC~4104 illustrate the fact that the BGGs of some FGs may show emission lines in their spectra. However, only a small percentage of BGGs are blue and star-forming, and so potentially exhibiting star-formation induced H$\alpha$ lines. The others, as NGC~4104, show emission lines originating from their centre and mainly due to their AGN activity. In NGC~4104, this AGN was probably reactivated in a recent past by some merging activity, therefore questioning the passive status of the central regions of FGs.

In conclusion, BGGs of FGs and non-FGs are found to differ morphologically, suggesting they have had somewhat different formation histories, but not sufficiently to make their stellar populations significantly different.

As a parallel project to increase the number of known FGs, we are presently confirming spectroscopically the subsample of FG candidates detected in the CFHTLS survey with a high probability by \cite{Adami+20}. This will be the topic of a future paper.

\begin{acknowledgements}
We thank the referee, R\'emi Cabanac, for his careful and constructive reading of the manuscript.
We are grateful to Nicolas Clerc for matching our FG and non-FG catalogues with his XCLASS X-ray catalogue. 
F.D. acknowledges support from CNES. 
I.M. acknowledges financial support from the State Agency for Research of the Spanish MCIU, through the "Center of Excellence Severo Ochoa" award to the Instituto de Astrofísica de Andalucía (SEV-2017-0709), and through PID2019-106027GB-C41. F.S. acknowledges support from a CNES Postdoctoral Fellowship.
 
This work is based on data obtained as part of UNIONS (initially CFIS), using data from a CFHT large program of the National Research Council of Canada and the French Centre National de la Recherche Scientifique. Based on observations obtained with MegaPrime/MegaCam, a joint project of CFHT and CEA Saclay, at the Canada-France-Hawaii Telescope (CFHT) which is operated
by the National Research Council (NRC) of Canada, the Institut National des Science de l’Univers (INSU) of the Centre National de la Recherche Scientifique (CNRS) of France, and the University of Hawaii.
Pan-STARRS is a project of the Institute for Astronomy of the University of Hawaii, and is supported by the NASA SSO Near Earth Observation Program under grants 80NSSC18K0971, NNX14AM74G, NNX12AR65G, NNX13AQ47G, NNX08AR22G, and by the State of Hawaii.
 
Funding for the Sloan Digital Sky Survey IV has been provided by the Alfred P. Sloan Foundation, the U.S.  Department of Energy Office of Science, and the Participating Institutions. SDSS-IV acknowledges support and resources from the Center for High-Performance Computing at the University of Utah. The SDSS web site is www.sdss.org.

SDSS-IV is managed by the Astrophysical Research Consortium for the  Participating Institutions of the SDSS Collaboration including the Brazilian Participation Group, the Carnegie Institution for Science, Carnegie Mellon University, the Chilean Participation Group, the French Participation Group, Harvard-Smithsonian Center for Astrophysics.

Based in part on observations made at Observatoire de Haute Provence (CNRS), France.

This research has made use of the MISTRAL database, based on observations made at Observatoire de Haute Provence (CNRS), France, with the MISTRAL 
spectro-imager, and operated at CeSAM (LAM), Marseille, France.

Based on observations collected at Centro Astronomico Hispano en Andalucia (CAHA) at Calar Alto, operated jointly by Instituto de Astrofisica de Andalucia (CSIC) and Junta de Andalucia.

Based on observations taken with the Nordic Optical Telescope on La Palma (Spain).

\end{acknowledgements}

\bibliographystyle{aa}
\bibliography{sample.bib}

\begin{appendix}

\section{Lists of FG candidates and non-FGs}

The list of the 87 FG candidates is given in Table~\ref{tab:88FG} with the principal properties of their BGG. The list of the 30 non-FGs for which the morphological properties were analysed are shown in Table~\ref{tab:30NonFG}.
The full list of 100 non-FGs will be available in electronic form at the CDS, together with the list of 87 FG candidates.

\begin{table*}[t]
\centering
\centering
\setcounter{table}{0}
\caption{Sample of 87 FG candidates. The columns are: running number, J2000 right ascension and declination (in degrees), spectroscopic redshift, logarithm of the BGG mass, absolute magnitudes in the g and r bands, virial radius, logarithm of the halo mass, UNIONS data when available with relevant photometric band(s).}
\label{tab:88FG}
\begin{tabular}{rrrrrrrrrl}
\hline \hline
Number & RA & DEC & z & logM$_{\rm BGG}$ & Mabs$_g$ & Mabs$_r$ & R$_{virial}$ (Mpc) & logM$_{halo}$ & UNIONS \\ 
\hline
1       &  4.6490 &-10.5378 &0.1463 &11.521 &-21.358 &-22.369 & 0.6069 & 13.053 & \\             
2       & 10.6416 & -9.9117 &0.0585 &11.194 &-21.325 &-22.327 & 0.4548 & 13.436 & \\             
3       & 22.3493 & 15.4461 &0.1727 &11.279 &-19.754 &-20.503 & 0.6696 & 13.628 & u \\           
4       & 24.9841 & -9.2424 &0.0421 &11.807 &-20.801 &-21.759 & 0.4630 & 13.109 & \\             
5       & 26.3686 &-10.0933 &0.0550 &11.493 &-20.868 &-21.811 & 0.7318 & 13.131 & \\             
6       & 58.2920 & -5.4971 &0.1230 &11.525 &-21.666 &-22.587 & 0.7250 & 13.055 & \\             
7       &117.6895 & 17.1722 &0.0727 &11.586 &-20.272 &-21.221 & 0.4433 & 13.020 & \\             
8       &118.8856 & 27.7361 &0.0748 &11.721 &-21.282 &-22.286 & 0.5019 & 13.672 & u \\           
9       &120.0094 & 51.6025 &0.0823 &11.644 &-20.168 &-21.120 & 0.5096 & 13.054 & r rLSB\\       
10      &124.8362 & 20.2687 &0.0816 &11.753 &-21.506 &-22.476 & 0.4569 & 13.070 & u \\           
11      &128.8010 & 31.7042 &0.0684 &11.424 &-21.369 &-22.295 & 0.4432 & 13.689 & u r rLSB \\    
12      &131.0486 & 23.5347 &0.0768 &11.464 &-21.091 &-22.050 & 0.4653 & 13.577 & u \\           
13      &131.1130 & 53.4878 &0.0616 &11.408 &-21.085 &-21.925 & 0.4823 & 13.467 & \\             
14      &131.7862 & 19.6311 &0.0312 &11.025 &-22.859 &-21.569 & 0.4403 & 13.607 & u \\           
15      &132.5318 &  2.6479 &0.0597 &11.562 &-20.921 &-21.866 & 0.6485 & 13.517 & \\             
16      &136.8240 & 16.7384 &0.0522 &11.389 &-20.574 &-21.552 & 0.4463 & 13.160 & \\             
17      &137.5553 & 38.7321 &0.0978 &11.531 &-20.399 &-21.346 & 0.6854 & 13.198 & u r rLSB\\     
18      &139.7271 & 50.0207 &0.0343 &11.495 &-20.761 &-21.739 & 0.4935 & 13.150 & \\             
19      &141.3023 &  5.3517 &0.0760 &11.658 &-20.682 &-21.606 & 0.5196 & 13.121 & \\             
20      &142.1698 & 12.6173 &0.0282 &10.986 &-21.118 &-22.104 & 0.4655 & 13.133 & \\             
21      &144.5160 & 42.9743 &0.0468 &11.207 &-20.680 &-21.613 & 0.4771 & 13.102 & u r rLSB \\    
22      &150.5563 & 11.3197 &0.0550 &11.284 &-19.868 &-20.831 & 0.4730 & 13.101 & \\             
23      &150.7645 & 16.6710 &0.0706 &11.591 &-21.217 &-22.207 & 0.4664 & 13.063 & \\             
24      &151.9548 & 39.7382 &0.0808 &11.621 &-21.008 &-21.952 & 0.5842 & 13.186 & r rLSB \\      
25      &154.8047 & 15.0122 &0.0815 &11.512 &-20.524 &-21.498 & 0.6847 & 13.007 & \\             
26      &155.9830 &  7.9813 &0.1033 &11.604 &-21.229 &-22.230 & 0.5261 & 13.189 & \\             
27      &157.3174 & 15.4637 &0.0570 &11.261 &-21.377 &-22.259 & 0.4406 & 13.605 & \\             
28      &161.1391 & 45.5408 &0.1122 &11.577 &-20.049 &-21.001 & 0.5509 & 13.043 & r rLSB \\      
29      &162.1012 &  5.2697 &0.0699 &11.235 &-20.415 &-21.334 & 0.5064 & 13.111 & \\             
30      &164.5891 &  3.6572 &0.0567 &11.312 &-21.166 &-22.145 & 0.4939 & 13.235 & \\             
31      &165.8158 & 54.1113 &0.0703 &11.510 &-21.037 &-21.990 & 0.5013 & 13.263 & r rLSB \\      
32      &175.0114 & 15.7188 &0.0844 &11.483 &-20.261 &-21.208 & 0.4708 & 13.156 & \\             
33      &175.5136 &  3.0047 &0.0406 &11.151 &-20.327 &-21.209 & 0.4457 & 13.210 & u \\           
34      &178.8740 & 49.7966 &0.0535 &11.139 &-20.907 &-21.818 & 0.4519 & 13.180 & u \\           
35      &179.8952 & 40.6662 &0.0666 &11.425 &-20.782 &-21.727 & 0.4915 & 13.306 & u r\\          
36      &180.1458 & 32.6646 &0.0715 &11.463 &-20.595 &-21.566 & 0.5057 & 13.328 & u r rLSB \\    
37      &181.2686 & 40.7910 &0.0525 &11.050 &-20.526 &-21.520 & 0.4447 & 13.210 & u r \\         
38      &181.3191 & 21.0103 &0.0746 &11.565 &-21.133 &-22.111 & 0.5507 & 13.100 & u \\           
39      &181.4802 & 25.2690 &0.1007 &11.722 &-20.689 &-21.624 & 0.5222 & 13.024 & u \\           
40      &181.5380 & -2.9481 &0.0256 &10.847 &-21.123 &-22.091 & 0.7142 & 13.389 & \\             
41      &182.3114 & 67.6405 &0.0599 &11.234 &-20.413 &-21.392 & 0.4919 & 13.103 & \\             
42      &183.1793 & 61.9709 &0.0496 &11.323 &-20.877 &-21.818 & 0.4802 & 13.110 & \\             
43      &184.6847 & 44.7812 &0.0383 &11.233 &-21.255 &-22.150 & 0.4753 & 13.326 & r rLSB\\       
44      &192.5297 & 52.8502 &0.0330 &11.385 &-20.683 &-21.649 & 0.4422 & 13.063 & \\             
45      &198.8502 &  7.8469 &0.0926 &11.614 &-21.053 &-21.931 & 0.5359 & 13.727 & \\             
46      &201.2748 &  6.3122 &0.0825 &11.437 &-20.423 &-21.364 & 0.4478 & 13.249 & \\             
47      &202.5798 & 11.5118 &0.0377 &11.184 &-21.247 &-22.226 & 0.5317 & 13.405 & \\             
48      &203.7632 & 35.4873 &0.0635 &11.011 &-20.667 &-21.559 & 0.4752 & 13.103 & u \\           
49      &204.6871 & 15.4295 &0.0745 &11.343 &-21.080 &-22.011 & 0.5456 & 13.027 & \\             
50      &205.0684 & 56.5015 &0.0998 &11.580 &-21.328 &-22.321 & 0.7522 & 13.499 & r \\           
51      &207.4004 & 28.4404 &0.0746 &11.404 &-21.001 &-22.005 & 0.4925 & 13.282 & u \\           
52      &210.5004 & 45.5619 &0.0654 &11.477 &-20.185 &-21.081 & 0.4571 & 13.022 & r \\           
53      &211.7204 & -1.7302 &0.0700 &11.443 &-20.505 &-21.461 & 0.4704 & 13.145 & \\             
54      &216.6863 &  9.1793 &0.0550 &11.335 &-20.687 &-21.649 & 0.4883 & 13.044 & \\             
55      &218.8973 & 50.1900 &0.0691 &11.456 &-20.003 &-20.859 & 0.4993 & 13.021 & r rLSB \\      
56      &219.1753 &  9.9292 &0.0586 &11.726 &-21.873 &-22.841 & 0.6403 & 13.901 & \\             
57      &219.6765 & 30.4659 &0.0707 &11.648 &-20.980 &-21.876 & 0.4580 & 13.201 & u \\           
58      &223.2020 & 32.3799 &0.0878 &11.873 &-21.724 &-22.760 & 0.7595 & 14.007 & u \\           
59      &225.6145 & 19.7352 &0.0972 &11.550 &-19.964 &-20.867 & 0.6811 & 13.002& \\
\hline
\end{tabular}
\end{table*}

\begin{table*}[ht]
\centering
\setcounter{table}{0}
  \caption{Continued.}
\begin{tabular}{rrrrrrrrrl}
  \hline
  \hline
  Number & RA & DEC & z & logM$_{\rm BGG}$ & Mabs$_g$ & Mabs$_r$ & R$_{virial}$ (Mpc) & logM$_{halo}$ \\ 
\hline
60      &225.8131 & 36.1477 &0.0733 &11.649 &-21.754 &-22.699 & 0.5928 & 13.734 & u \\              
61      &226.1660 & 53.8232 &0.0379 &11.403 &-20.567 &-21.559 & 0.5158 & 13.133 & r \\              
62      &227.3765 & 46.4927 &0.0378 &11.195 &-19.350 &-20.342 & 0.5162 & 13.072 & r\\               
63      &227.4669 & -0.3847 &0.0711 &11.775 &-20.964 &-21.874 & 0.4527 & 13.129 & \\                
64      &228.2578 & 28.4928 &0.0786 &11.390 &-20.520 &-21.464 & 0.4803 & 13.037 & u \\              
65      &228.7253 & 42.0131 &0.1348 &11.968 &-20.222 &-21.162 & 0.9219 & 13.185 & u r rLSB\\        
66      &232.6162 & -0.2305 &0.0869 &11.774 &-20.435 &-21.238 & 0.6042 & 13.090\\                   
67      &233.3931 & 33.6996 &0.0677 &11.502 &-21.141 &-22.153 & 0.4443 & 13.212 & u r rLSB \\       
68      &233.4166 & 24.4047 &0.0434 &11.335 &-20.677 &-21.642 & 0.4801 & 13.158 & u \\              
69      &234.9934 & 48.5938 &0.0677 &11.391 &-20.930 &-21.872 & 0.5596 & 13.091 & r rLSB \\         
70      &235.6059 & 13.9567 &0.0926 &11.635 &-20.781 &-21.717 & 0.4694 & 13.026 & \\                
71      &236.6317 & 12.1427 &0.0720 &11.800 &-20.121 &-21.008 & 0.6834 & 13.113 & \\                
72      &237.0568 & 29.9150 &0.0960 &11.500 &-21.384 &-22.344 & 0.5031 & 13.069 & u r rLSB\\        
74      &239.8565 & 42.2635 &0.0605 &11.420 &-20.906 &-21.841 & 0.5515 & 13.068 & r rLSB \\         
75      &242.3237 &  4.0439 &0.0551 &10.939 &-20.982 &-21.903 & 0.4439 & 13.197 & \\                
76      &246.6880 & 24.1490 &0.0589 &11.117 &-20.538 &-21.497 & 0.4370 & 13.164 & u \\              
77      &251.8154 & 33.9324 &0.0673 &11.622 &-20.905 &-21.875 & 0.4699 & 13.596 & r rLSB \\         
78      &252.4850 & 35.2121 &0.0996 &11.589 &-19.763 &-20.775 & 0.4768 & 13.000 & u r rLSB\\        
79      &255.2097 & 23.0110 &0.0094 &11.906 &-21.262 &-22.206 & 0.8850 & 13.252 & u \\              
80      &255.8944 & 58.9176 &0.0762 &11.516 &-20.669 &-21.656 & 0.4717 & 13.026 & u r rLSB \\       
81      &317.6327 &  0.8951 &0.0681 &11.319 &-20.853 &-21.800 & 0.4475 & 13.109 & u \\              
82      &322.3274 & 11.2010 &0.0890 &11.755 &-21.004 &-21.990 & 0.4537 & 13.310 & u \\              
83      &327.4432 & -7.4425 &0.0903 &11.582 &-21.746 &-22.652 & 0.6729 & 13.590 & u \\              
84      &336.5609 &  0.6678 &0.0364 &11.016 &-20.888 &-21.877 & 0.4711 & 13.091 & u \\              
85      &340.9873 &  1.0005 &0.0580 &11.195 &-20.881 &-21.756 & 0.5032 & 13.297 & u \\              
86      &342.3251 & 12.6305 &0.1337 &10.374 &-19.661 &-20.730 & 0.4602 & 13.072 & u \\              
87      &346.9538 &  0.9405 &0.0418 &11.459 &-21.107 &-22.059 & 0.4846 & 13.596 & u \\              
88      &348.8094 & -1.2422 &0.0251 &10.848 &-20.366 &-21.327 & 0.4662 & 13.012 & \\                
\hline
\end{tabular}
\end{table*}

\begin{table*}
\centering
\centering
\caption{Control sample of 30 non-FG candidates. The columns are: name, J2000 right ascension and declination (in degrees), spectroscopic redshift, logarithm of the BGG mass, absolute magnitudes in the g and r bands, virial radius, logarithm of the halo mass. The availability of UNIONS data is not indicated, since they all have UNIONS data in the u, r, and rLSB bands.}
\label{tab:30NonFG}
\begin{tabular}{rrrrrrrrr}
\hline \hline
Name & RA & DEC & z & logM$_{\rm BGG}$ & Mabs$_g$ & Mabs$_r$ & R$_{Virial}$ (Mpc) & logM$_{halo}$ \\ 
\hline
NonFG6  &121.2352 & 31.2595 &0.0732 &11.363 &-20.550 &-21.495 & 0.4882 & 13.152 \\
NonFG8  &121.5296 & 39.3424 &0.0645 &11.480 &-19.942 &-20.897 & 0.5038 & 13.190 \\
NonFG9  &121.6227 & 33.2801 &0.0838 &10.829 &-20.221 &-21.072 & 0.4785 & 13.131 \\
NonFG12 &123.3757 & 30.3811 &0.0753 &11.197 &-19.908 &-20.894 & 0.4397 & 13.017 \\
NonFG15 &124.1161 & 33.3768 &0.1084 &11.150 &-20.659 &-21.640 & 0.6743 & 13.588 \\
NonFG16 &124.7523 & 34.9870 &0.0623 &11.159 &-20.596 &-21.592 & 0.5076 & 13.198 \\
NonFG17 &125.0182 & 41.3390 &0.1021 &11.216 &-20.396 &-21.356 & 0.5278 & 13.266 \\
NonFG19 &126.1656 & 32.3262 &0.0683 &11.357 &-20.572 &-21.526 & 0.5401 & 13.282 \\
NonFG22 &126.5740 & 29.5435 &0.1100 &11.251 &-20.850 &-21.837 & 0.5273 & 13.268 \\
NonFG23 &126.8353 & 34.2165 &0.0876 &10.973 &-20.556 &-21.531 & 0.5129 & 13.223 \\
NonFG24 &127.3892 & 31.6653 &0.0900 &11.595 &-20.422 &-21.453 & 0.5514 & 13.318 \\
NonFG29 &128.2929 & 30.4762 &0.1075 &11.301 &-20.196 &-21.128 & 0.4921 & 13.177 \\
NonFG30 &128.3209 & 30.9804 &0.0936 &11.243 &-19.755 &-20.676 & 0.4963 & 13.182 \\
NonFG31 &128.9864 & 30.1257 &0.0933 &10.824 &-20.749 &-21.711 & 0.4650 & 13.097 \\
NonFG33 &129.4784 & 36.9984 &0.0548 &11.574 &-20.437 &-21.297 & 0.4853 & 13.137 \\
NonFG34 &129.5675 & 35.0896 &0.0647 &11.434 &-20.314 &-21.331 & 0.4675 & 13.092 \\
NonFG48 &131.7605 & 31.4298 &0.0660 &11.346 &-19.842 &-20.664 & 0.6012 & 13.420 \\
NonFG56 &133.3498 & 37.3566 &0.1036 &11.083 &-21.203 &-22.141 & 0.5211 & 13.250 \\
NonFG58 &133.6843 & 35.5656 &0.0882 &11.258 &-21.218 &-22.227 & 0.7549 & 13.726 \\
NonFG63 &134.0557 & 37.7081 &0.0938 &10.904 &-21.090 &-22.034 & 0.4891 & 13.163 \\
NonFG67 &135.3629 & 34.4643 &0.0655 &11.307 &-20.122 &-21.080 & 0.5484 & 13.301 \\
NonFG69 &136.0299 & 30.8830 &0.0635 &11.212 &-20.990 &-21.967 & 0.5083 & 13.201 \\
NonFG78 &140.8148 & 33.5107 &0.0424 &10.670 &-20.200 &-21.173 & 0.4663 & 13.080 \\
NonFG82 &142.0339 & 35.8988 &0.1108 &11.454 &-20.930 &-21.906 & 0.4497 & 13.061 \\
NonFG84 &143.2806 & 32.2718 &0.0735 &10.961 &-21.084 &-22.079 & 0.5072 & 13.202 \\
NonFG86 &143.7315 & 32.8416 &0.0610 &11.287 &-19.442 &-20.364 & 0.5401 & 13.279 \\
NonFG90 &144.2122 & 30.2841 &0.1111 &11.305 &-21.027 &-21.855 & 0.4677 & 13.112 \\
NonFG95 &148.1868 & 40.7628 &0.0925 &11.150 &-21.372 &-22.324 & 0.4461 & 13.043 \\
NonFG96 &148.3277 & 34.7647 &0.0502 &11.332 &-20.690 &-21.682 & 0.4690 & 13.090 \\
NonFG98 &148.5450 & 32.4480 &0.0869 &11.200 &-20.812 &-21.764 & 0.6378 & 13.506 \\
\hline
\end{tabular}
\end{table*}

\section{Spectra of NGC~4104}

\begin{figure*}[h]
\centering
\includegraphics[width=17.cm,angle=0]{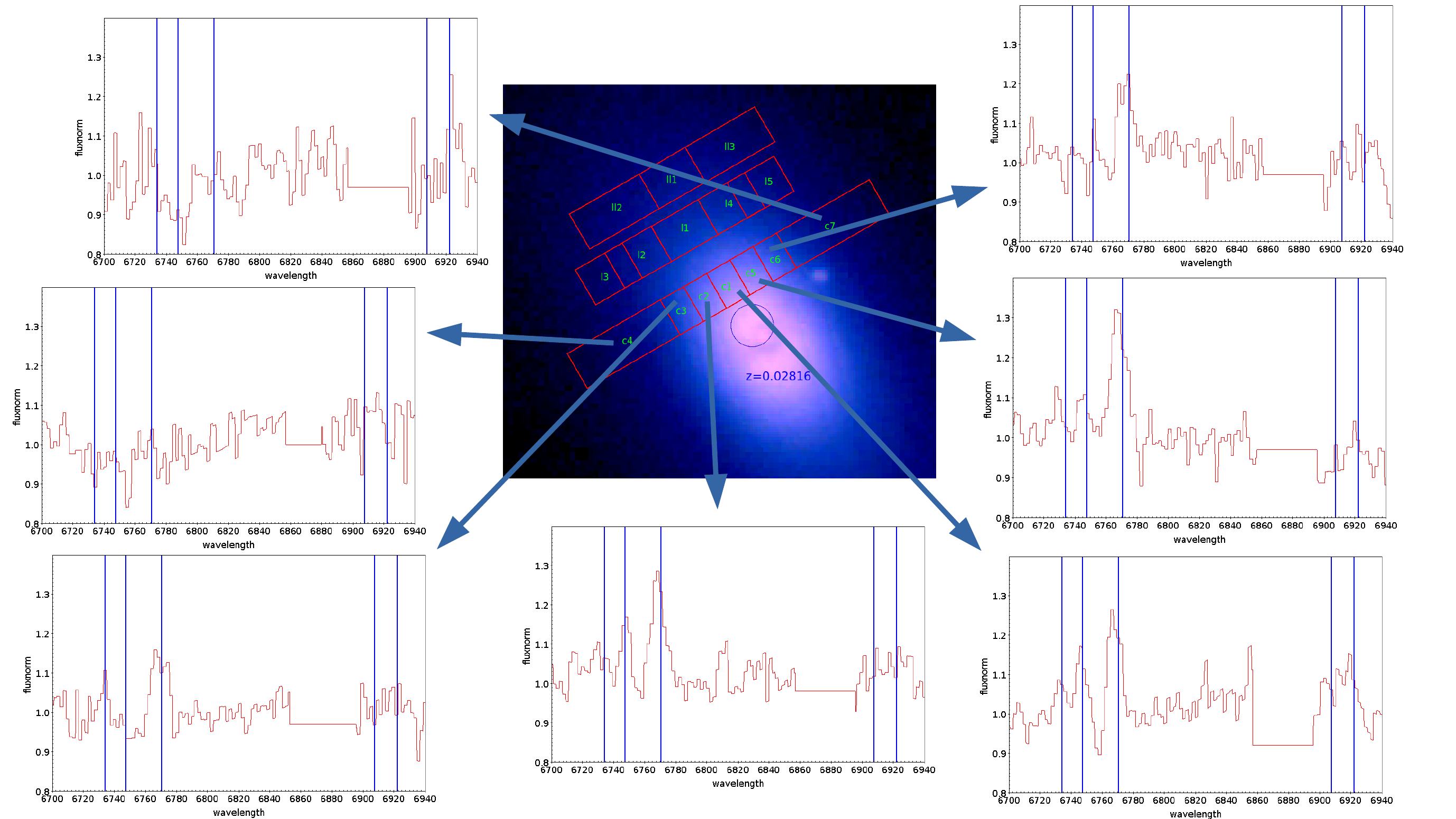}
\includegraphics[width=17.cm,angle=0]{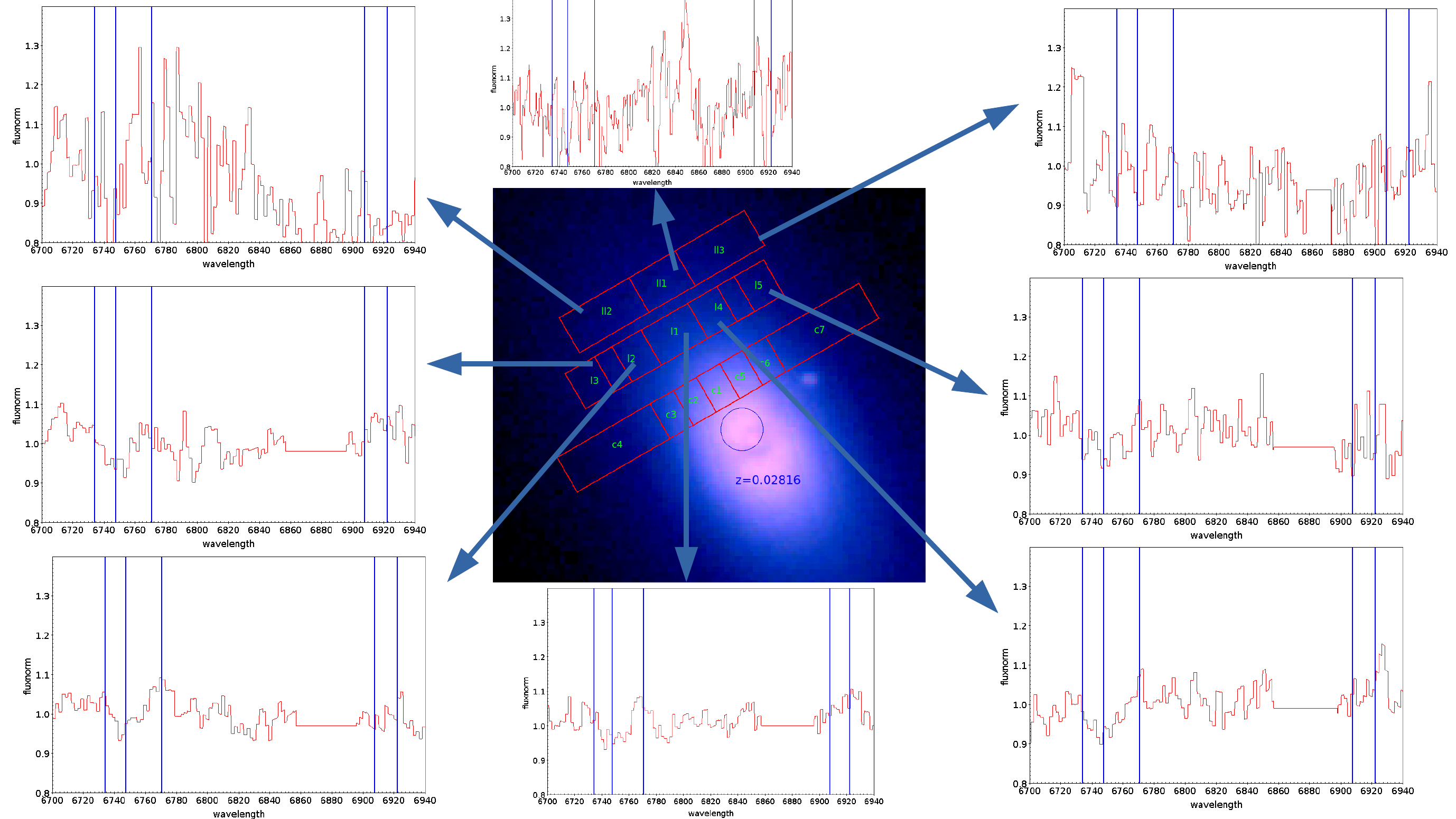}
\caption{MISTRAL spectroscopic observations of 15
  different regions of NGC~4104. Spectra are normalized in flux.  The
  location of the SDSS fiber is shown as the blue central circle.
  Wavelength domain is limited to [6700, 6940]\AA . }
  \label{fig:spectres_zoom}
\end{figure*}

Illustrations of the observations obtained with the MISTRAL spectrograph at Observatoire de Haute Provence are given in Fig.~\ref{fig:spectres_zoom}.

\end{appendix}
\end{document}